\documentclass{ws-ijmpd}
\usepackage[T1]{fontenc}
\usepackage{microtype}
\usepackage{amsmath,amssymb,amsfonts,mathtools}
\usepackage{mathrsfs}
\DeclareFontFamily{U}{rsfs}{\skewchar\font127}
\DeclareFontShape{U}{rsfs}{m}{n}{<-> rsfs10}{}
\usepackage{bm}

\usepackage{graphicx}
\usepackage{xurl}
\usepackage{booktabs}
\usepackage{xcolor}
\usepackage[superscript]{cite}
\usepackage[hidelinks]{hyperref}

\newcommand{\dd}{\mathrm{d}}

\begin{document}
\raggedbottom

\hypersetup{
  pdftitle={Apparent Horizon Thermodynamics in an Exponential f(Q) Gravity Model},
  pdfauthor={A. Oliveros; Ivan R. Vasquez}
}

\title{Apparent horizon thermodynamics in an exponential $f(Q)$ gravity model}
\markboth{A. Oliveros and Ivan R. Vasquez}
{Apparent horizon thermodynamics in exponential $f(Q)$ gravity}

\author{A. Oliveros$^{\dagger}$, Ivan R. Vasquez$^{\ast}$}
\address{Grupo de Física de Partículas Elementales y Cosmología, Programa de Física,
Universidad del Atlántico, Carrera 30 No. 8-49,
Puerto Colombia, Atl\'antico, Colombia\\
$^{\dagger}$Corresponding author: \email{alexanderoliveros@mail.uniatlantico.edu.co}\\
$^{\ast}$\email{itsivanvasquez@gmail.com}}

\maketitle

\begin{abstract}
We investigate the thermodynamics of the apparent horizon in an exponential
$f(Q)$ gravity model characterized by the two parameters $b$ and $n$, within a
spatially flat Friedmann--Lema\^itre--Robertson--Walker background and the
coincident gauge. Focusing on the $n=1$ solution, we use
the approximate cosmological solution for the Hubble parameter up to second
order in the exponential parameter $b$ to study the redshift evolution of the
apparent-horizon radius and the Kodama--Hayward temperature. We formulate
Hayward's unified first law in terms of an effective form of the Misner--Sharp--Hernandez mass. Aswell as the
work density, and the energy-supply vector, and show that the horizon dynamics
admits an equilibrium thermodynamic description. The associated entropy
differential is proportional to $f_Q+2Qf_{QQ}$, yielding exponentially
suppressed corrections to the Bekenstein--Hawking area law and recovering
$S=A/4$ in the limit $b\to0$. We then examine the generalized second law (GSL) by
including the entropy of matter inside the apparent horizon. When the matter and
horizon temperatures are identified, we adopt the GSL viability criterion
$f_Q+2Qf_{QQ}\geq0$ previously derived in the literature. The observationally
motivated best-fit values of $b$ satisfy this condition over the redshift
interval studied and produce
departures from $\Lambda$CDM mainly at late times. In contrast, sufficiently
large positive values, approximately $b>0.26$, can violate the GSL in the future
region $z<0$. These results show that apparent-horizon
thermodynamics provides a complementary constraint on the parameter space of
exponential $f(Q)$ cosmology.
\end{abstract}

\section{Introduction}
The discovery of the late-time accelerated expansion of the Universe, first
established through observations of Type Ia supernovae by the High-Z Supernova
Search Team and the Supernova Cosmology Project
\cite{Riess:1998cb,Perlmutter:1998np}, has motivated the search for
gravitational scenarios that can reproduce the observed cosmic dynamics without
relying exclusively on a cosmological constant or on an unspecified dark-energy (DE)
component. Although the $\Lambda$CDM model provides an economical
phenomenological description, it faces both theoretical difficulties, such as
the fine-tuning and coincidence problems
\cite{Weinberg:1988cp,Sahni:1999gb,Peebles:2002gy,Copeland:2006wr}, and
observational tensions. A prominent example is the current tension in the Hubble
constant $H_0$. The value
inferred from the CMB within the standard model differs
from local distance-ladder determinations
\cite{Planck:2018vyg,Riess:2021jrx}. More recently, DESI Baryon-Acoustic-
Oscillation measurements, especially when combined with CMB and Type Ia
supernova data, have also suggested that the DE equation of state may
be time dependent rather than strictly described by a cosmological constant
\cite{DESI:2024mwx,DESI:2025zgx}. These open issues continue to encourage
the study of extensions of General Relativity (GR) in which the gravitational sector
itself contributes effectively to the accelerated regime of expansion. In this context, modified gravity theories based on the geometric structure of spacetime have
become a central arena for connecting cosmology, fundamental interactions, and
thermodynamics. For broad reviews of modified gravity and its cosmological
applications, see Refs.~\cite{Nojiri:2010wj,Clifton:2011jh,Joyce:2016vqv}.

Symmetric teleparallel gravity offers a particularly interesting formulation of
gravitation. In contrast with the standard curvature description of GR, and with torsion-based teleparallel gravity, this framework assigns
the gravitational interaction to non-metricity while imposing vanishing curvature
and torsion. The equivalent formulation known as Coincident General Relativity (CGR)
was developed in Refs.~\cite{Jimenez:2018bfs,Jimenez:2019ovq}, where the role of
the non-metricity scalar $Q$ was emphasized as part of the geometrical trinity
of gravity. A natural extension is then obtained by replacing $Q$ with a
general functional $f(Q)$, in close analogy with $f(R)$ and $f(T)$ theories.
Such models have attracted considerable attention because they can modify the
background expansion and the effective gravitational coupling while retaining a
comparatively simple second-order structure in the cosmological field equations.
This second-order structure, however, does not by itself determine the complete
dynamical content of the theory. Perturbative analyses have identified
additional scalar modes that may disappear on highly symmetric backgrounds,
raising the possibility of strong coupling. Moreover, Hamiltonian and covariant
perturbation studies have obtained different degree-of-freedom counts and have
shown that the propagating spectrum can depend on the background and on the
choice of affine connection; some cosmological branches have also been found to
contain ghost or strongly coupled modes
\cite{Jimenez:2019qfb,DAmbrosio:2023cwr,Tomonari:2023wcs,Gomes:2023tur,Heisenberg:2023lgi}.
Consequently, the precise dynamical degrees of freedom of generic $f(Q)$ models
and their stability has been addressed and discussed in \cite{Gomes:2023tur, BeltranJimenez:2026wja}, establishing a pathological character of $f(Q)$ theories, although its use in literature still stimulate discussions and results.
Observational analyses of $f(Q)$ cosmologies have shown that non-metricity
corrections can be compatible with late-time data and may provide viable
alternatives to the standard scenario \cite{Lazkoz:2019sjl,Anagnostopoulos:2021ydo}.
For further details on the foundations, applications, and current status of
$f(Q)$ theories, see the recent review in Ref.~\cite{Heisenberg:2023lru}.

Among the possible choices for the gravitational Lagrangian, exponential
corrections are especially useful because they can be arranged to be relevant at
cosmological scales while recovering the GR limit in suitable
regimes. Exponential functions have been widely employed in modified gravity and
DE model building, where they often generate smooth transitions between
matter domination and accelerated expansion. In curvature-based $f(R)$ gravity,
exponential models have been proposed both to account for late-time cosmic
acceleration and to construct nonsingular scenarios unifying inflation with the
current accelerated era \cite{Linder:2009jz,Elizalde:2010ts}. Exponential
corrections have also been investigated in torsion-based $f(T)$ gravity, for
example in studies of the existence and stability of Einstein-static solutions
\cite{Li:2013xea}. In $f(Q)$ gravity, an exponential model modifies the
dependence of the effective gravitational coupling on the non-metricity scalar
and therefore may affect not only the expansion history but also the
thermodynamic properties associated with cosmological horizons. Recent studies
of exponential-type $f(Q)$ models have explored their background cosmological
dynamics, observational viability, energy conditions, matter perturbations,
large-scale-structure signatures, and phase-space structure
\cite{Khyllep:2022spx,Narawade:2023jfl,Sokoliuk:2023eha,Mhamdi:2024glb,Oliveros:2023bdv,Vasquez:2025ywd,Vasquez:2025tbg,Ganz:2025ydt, Anagnostopoulos:2021ydo, Odintsov:2017qif}.

The thermodynamic interpretation of gravitational dynamics provides an
independent criterion for testing the consistency of such models. The area law
for black-hole entropy \cite{Bekenstein:1973ur,Hawking:1975vcx}, in general the so-called laws of black hole mechanics \cite{Bardeen:1973gs} and Jacobson's 
derivation of the Einstein equations from the Clausius relation
\cite{Jacobson:1995ab} suggest a deep relation between gravity and
thermodynamics. In cosmology, the apparent horizon plays a distinguished role:
for Friedmann--Lema\^itre--Robertson--Walker spacetimes, the Friedmann equations
can be written in a form analogous to the first law of thermodynamics at this
horizon \cite{Cai:2005ra,Akbar:2006kj}. This connection has been generalized to
several modified gravity theories, where the horizon entropy, effective Newton
constant, and possible entropy production terms encode deviations from the
Einstein case \cite{Bamba:2009gq,Bamba:2012rv}. Recent studies have also begun
to examine the apparent-horizon thermodynamics and GSL in
non-metricity-based cosmology \cite{Rao:2024rhn}, as well as correspondences between modified gravity and entropic cosmology \cite{Nojiri:2025fiu, Nojiri:2025gkq}. Related recent analyses have
extended this thermodynamic program to $f(Q,\mathcal{T})$ gravity, where the
trace of the energy--momentum tensor contributes to the entropy balance, and to
torsion-free $f(Q)$ scenarios in which dynamical diagnostics are combined with
tests of the GSL, Gibbs free energy, and heat capacity
\cite{Pradhan:2025gsl,Pervaiz:2026npb}.

The aim of this work is to analyze apparent-horizon thermodynamics in a specific 
exponential $f(Q)$ gravity model, which was introduced by one of the authors in \cite{Oliveros:2023bdv}. We focus on the form of the horizon entropy,
the validity of the first law, and the GSL including the
entropy of matter inside the apparent horizon. This analysis is intended to
identify the thermodynamic conditions under which the exponential model remains
viable and to clarify how the parameters of the non-metricity correction enter
the entropy balance during cosmological evolution. In particular, for the
$n=1$ branch, we investigate the thermodynamic constraints imposed by the GSL
on the allowed values of $b$.

The paper is organized as follows. In Sec.~\ref{sec:model} we summarize the
basic elements of symmetric teleparallel gravity and introduce the exponential
$f(Q)$ model. In Sec.~\ref{sec:frwmetric-ah} we specialize the geometry to a
spatially flat FLRW background, identify the apparent horizon, and derive the
Kodama--Hayward temperature used in the thermodynamic analysis. In
Sec.~\ref{sec:thermodynamics} we formulate the thermodynamics of the apparent
horizon, including the first law and the GSL. Finally,
Sec.~\ref{sec:conclusion} contains our main conclusions and possible extensions.

\section{Symmetric Teleparallel Gravity and the Exponential Model}
\label{sec:model}

Symmetric teleparallel gravity is most naturally formulated within
metric--affine geometry, where the metric and affine connection are regarded as
independent structures. In the symmetric teleparallel sector one imposes
vanishing curvature and torsion,
\begin{equation}
  R^{\lambda}_{\phantom{\lambda}\rho\mu\nu}=0,
  \qquad
  T^{\lambda}_{\phantom{\lambda}\mu\nu}
  =2\Gamma^{\lambda}_{[\mu\nu]}=0,
  \label{eq:stg-constraints}
\end{equation}
so that non-metricity is the only nonvanishing geometric quantity associated
with the affine connection. This choice defines the non-metricity sector of the
geometrical trinity of gravity: the same gravitational interaction described by
curvature in GR can equivalently be represented, in this sector,
by the failure of the connection to preserve the metric under parallel
transport. Following the conventions used for the exponential model in
Ref.~\cite{Oliveros:2023bdv} and in the review~\cite{Heisenberg:2023lru}, the
non-metricity tensor is defined as
\begin{equation}
  Q_{\lambda\mu\nu} \equiv \nabla_{\lambda}g_{\mu\nu},\label{eq:nonmetricity-tensor}
\end{equation}
with the two independent traces
\begin{equation}
  Q_{\lambda} \equiv Q_{\lambda\phantom{\mu}\mu}^{\phantom{\lambda}\mu},
  \qquad
  \tilde Q_{\lambda} \equiv Q^{\mu}_{\phantom{\mu}\lambda\mu}.\label{eq:nonmetricity-traces}
\end{equation}
It is useful to introduce the disformation tensor
\begin{equation}
  L^{\lambda}_{\phantom{\lambda}\mu\nu}
  = -\frac{1}{2} g^{\lambda\sigma}
  \left(\nabla_{\nu}g_{\mu\sigma}
  + \nabla_{\mu}g_{\nu\sigma}
  - \nabla_{\sigma}g_{\mu\nu}\right),\label{eq:disformation-tensor}
\end{equation}
and the non-metricity conjugate
\begin{equation}
  P^{\lambda}_{\phantom{\lambda}\mu\nu}
  = -\frac{1}{2}L^{\lambda}_{\phantom{\lambda}\mu\nu}
  + \frac{1}{4}\left(Q^{\lambda}-\tilde Q^{\lambda}\right)g_{\mu\nu}
  - \frac{1}{4}\delta^{\lambda}_{\phantom{\lambda}(\mu}Q_{\nu)}.\label{eq:nonmetricity-conjugate}
\end{equation}
The non-metricity scalar is then constructed as
\begin{equation}
  Q = - Q_{\lambda\mu\nu}P^{\lambda\mu\nu}.\label{eq:nonmetricity-scalar}
\end{equation}

For the specific linear choice $f(Q)=Q$, this scalar reproduces the symmetric
teleparallel equivalent of GR up to a boundary contribution. The
nonlinear replacement $Q\mapsto f(Q)$ therefore plays a role analogous to the
extensions $R\mapsto f(R)$ and $T\mapsto f(T)$, but it modifies the
non-metricity sector while keeping the curvature and torsion constraints of
Eq.~\eqref{eq:stg-constraints}. We first present the covariant field equations
and then specialize them to the homogeneous and isotropic coincident-gauge branch
used throughout the cosmological and thermodynamic analysis.

The action for \(f(Q)\) gravity may be written as
\begin{equation}
  S = \int \dd^4 x\, \sqrt{-g}\left[\frac{1}{16\pi G} f(Q) + \mathcal{L}_m\right],
  \label{eq:fq-action}
\end{equation}
where $\mathcal{L}_m$ denotes the matter Lagrangian. Matter is assumed to be
minimally coupled to the metric, so that the energy--momentum tensor is defined
in the usual way from $\mathcal{L}_m$. Variation with respect to the metric
gives the field equations
\begin{equation}
  \frac{2}{\sqrt{-g}}\nabla_{\lambda}\left(\sqrt{-g}\, f_Q P^{\lambda}_{\phantom{\lambda}\mu\nu}\right)+\frac{1}{2}g_{\mu\nu}f+f_Q\left(P_{\mu\lambda\sigma}Q_{\nu}^{\phantom{\nu}\lambda\sigma}-2Q_{\lambda\sigma\mu}P^{\lambda\sigma}_{\phantom{\lambda\sigma}\nu}\right)=-8\pi G\,T_{\mu\nu},
  \label{eq:fq-field-equations}
\end{equation}
where $f_Q \equiv \dd f/\dd Q$ and $T_{\mu\nu}$ is the matter
energy--momentum tensor. As emphasized in the review~\cite{Heisenberg:2023lru},
these equations can also be recast as Einstein equations sourced by an effective
geometric sector. In that form, the combination $f-Qf_Q$ acts as an effective
cosmological contribution, while derivative terms proportional to $f_{QQ}$
encode genuine departures from the STEGR/GR sector. Therefore, the limit
$f_{QQ}=0$, with constant $f_Q$, recovers the Einstein dynamics up to a
possible cosmological constant and a rescaling of the gravitational coupling.
The independent variation with respect to the connection yields
\begin{equation}
  \nabla_{\mu}\nabla_{\nu}
  \left(\sqrt{-g}\, f_Q P^{\mu\nu}_{\phantom{\mu\nu}\lambda}\right)=0,\label{eq:connection-equation}
\end{equation}
Because the symmetric teleparallel connection is flat and torsionless, it is
locally a pure-gauge object. In the coincident gauge one may set
$\Gamma^{\lambda}_{\phantom{\lambda}\mu\nu}=0$, so that the non-metricity tensor
is determined entirely by metric derivatives. More general non-trivial flat and
torsionless connections may be retained in fully covariant treatments, but they
are outside the scope of the present work. Different connections and their use in cosmological models has been explored in \cite{Guzman:2024cwa}. For the homogeneous and isotropic
flat FLRW branch considered below, the connection equation is compatible with
the metric equations and matter conservation, allowing the background dynamics
to be studied directly through the modified Friedmann system. 

Hereafter we use geometrized units in which $G=1$. For a spatially flat FLRW geometry, the non-metricity scalar reduces to
\begin{equation}
  Q = 6H^2,\label{eq:q-flrw}
\end{equation}
where $H=\dot a/a$. Assuming a total energy density $\rho=\sum \rho_i$
and a total pressure $p=\sum p_i$, the metric field equations give the modified
Friedmann equations
\begin{align}
  6H^2 f_Q - \frac{1}{2}f &= 8\pi\rho,
  \label{eq:friedmann-1}\\
  \left(12H^2 f_{QQ}+f_Q\right)\dot H &= -4\pi(\rho+p),
  \label{eq:friedmann-2}
\end{align}
together with the usual conservation law
\begin{equation}
  \dot\rho + 3H(\rho+p)=0.\label{eq:continuity}
\end{equation}
These equations show explicitly how the derivatives of $f(Q)$ modify the
cosmological background while preserving the second-order character of the
background field equations. In particular, $f_Q$ rescales the effective
gravity sector, whereas $f_{QQ}$ controls the response of the theory to the
cosmological evolution of $Q=6H^2$.

Following Ref.~\cite{Oliveros:2023bdv}, for the exponential model we will consider a functional of the form:
\begin{equation}
  f(Q) = Q + 2\Lambda\, e^{-(b\Lambda/Q)^n},
  \label{fexp}
\end{equation}
where $b$ and $n$ are model parameters, and $\Lambda$ is the cosmological constant. Its first
and second derivatives, which enter the modified Friedmann equations, are
\begin{align}
  f_Q &= 1 + 2n\Lambda\,\frac{(b\Lambda)^n}{Q^{n+1}}
  \exp\left[-\left(\frac{b\Lambda}{Q}\right)^n\right],
  \label{eq:exponential-fq}
\end{align}
\begin{equation}
  f_{QQ} = 2n\Lambda\,\frac{(b\Lambda)^n}{Q^{n+2}}\left[n\left(\frac{b\Lambda}{Q}\right)^n-(n+1)\right]\times\exp\left[-\left(\frac{b\Lambda}{Q}\right)^n\right].
  \label{eq:exponential-fqq}
\end{equation}
Thus, in regimes where the exponential correction is suppressed, the model
approaches the GR limit $f(Q)\simeq Q$, whereas departures
from this limit can source an effective dark-energy contribution at late times.

\section{The FLRW metric and the apparent horizon}
\label{sec:frwmetric-ah}

In this section we specialize the cosmological background to a spatially flat
Friedmann--Lema\^i\ tre--Robertson--Walker spacetime and identify the apparent
horizon associated with the areal radius. This construction provides the
geometrical setting needed to define the Kodama vector, the Kodama--Hayward
surface gravity, and the horizon temperature that will enter the thermodynamic
analysis of the exponential $f(Q)$ model.

\begin{figure}[!ht]
  \centering
  \includegraphics[width=0.6\linewidth]{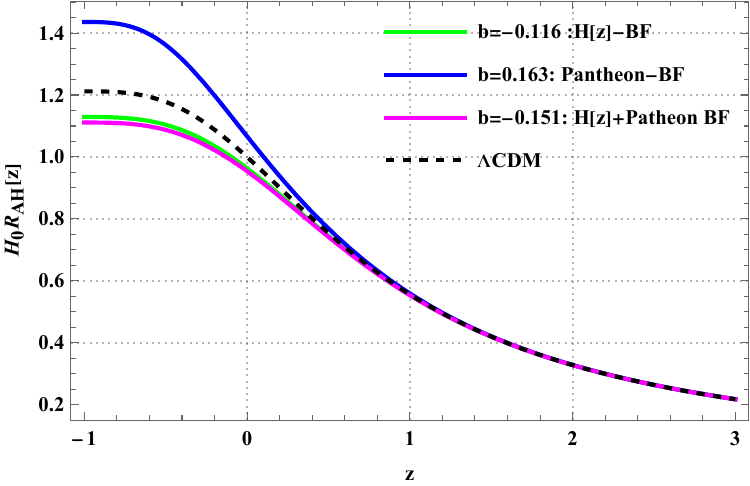}
  \caption{Redshift evolution of the dimensionless apparent-horizon radius
  $H_0R_{\rm AH}=H_0/H(z;b)$ in the exponential $f(Q)$ model. The curves
  correspond to the best-fit values $b=-0.116$ from $H(z)$ data, $b=0.163$
  from Pantheon data, and $b=-0.151$ from the joint $H(z)$+Pantheon analysis
  reported in Ref.~\cite{Oliveros:2023bdv}, together with the $\Lambda$CDM
  limit $b=0$. The numerical evaluation uses $\Omega_{m,0}=0.315$
  \cite{Planck:2018vyg}.}
  \label{fig:apparent-horizon-radius}
\end{figure}

The spatially flat FLRW metric in spherical coordinates is given by
\begin{equation}\label{metric1}
  \dd s^2 = -\dd t^2 + a^2(t)\left(\dd r^2 + r^2 \dd\Omega_{(2)}^2\right),
\end{equation}
where \(r\) is the comoving radial coordinate and $\dd\Omega_{(2)}^2 = \dd\theta^2 + \sin^2\theta\,\dd\phi^2$ is the line element on a unit two-sphere. The corresponding proper, or areal, radius is $R(t,r)=a(t)r$, whose differential is
\begin{align}
  \dd R = H R\,\dd t + a\,\dd r,
  \label{eq:areal-radius-differential}
\end{align}
where \(H=\dot{a}/a\). Now,
\begin{align}
    dr=\frac{1}{a}(dR-HR\, dt).
\end{align}
Replacing back in the metric \eqref{metric1} one gets:
\begin{align}
\label{metric2}
    ds^2 = -(1-H^2R^2)dt^2+dR^2-2HR\,dt\,dR+R^{2}d\Omega^{2}_{(2)}.
\end{align}
This form is usually called \textit{r-gauge}, or the metric said to be in Painlevé-Gullstrand form \cite{DiCriscienzo:2007pcr}. For the purpose of locating the apparent horizon, the crossed-term can be eliminated by introducing a new coordinate $T$, replacing the time coordinate, whose differential is given by:
\begin{equation}
    dT=\frac{1}{F(t,r)}(dt+\beta\, dR),
\end{equation}
such that $F(t,r)$ is an integrating factor where:
\begin{equation}
    \frac{\partial T}{\partial t}=\frac{1}{F}, \quad \frac{\partial T}{\partial R}=\frac{\beta}{F}.
\end{equation}
$\beta(t,r)$ is a function to be determined. In virtue of commutativity of second partial derivatives, the following condition must be satisfied to ensure $dT$ is an exact differential:
\begin{equation}
    \frac{\partial}{\partial R}\left(\frac{1}{F}\right)=\frac{\partial}{\partial t}\left(\frac{\beta}{F}\right).
\end{equation}
Thus, using $dt=FdT-\beta\, dR$ in \eqref{metric2}. Results in the following form:
\begin{equation}
\label{eq:metric-before-diagonalization}
\begin{split}
    ds^2={}&-(1-H^2R^2)F^2\,dT^2
    +2F[(1-H^2R^2)\beta-HR]dT\,dR\\
    &+[1-(1-H^2R^2)\beta^2+2\beta\,H\,R]dR^2
    +R^2\,d\Omega^{2}_{(2)}.
\end{split}
\end{equation}
Choosing $\beta(t,r)$ to be defined as,
\begin{align}
    \beta(t,r)=\frac{HR}{1-H^2R^2},
\end{align}
the crossed-term can be eliminated. The metric finally takes a Schwarzschild-like form in coordinates $(T,R,\theta,\phi)$ as:
\begin{equation}\label{metric3}
    ds^2=-(1-H^2R^2)F^2\,dT^2+\frac{dR^2}{1-H^2R^2}+R^2\,d\Omega^{2}_{(2)}.
\end{equation}
This Schwarzschild-like form of the metric is usually called Nolan-McVittie gauge, related to the embedding of a point mass in FLRW spacetime \cite{Nolan:1998xs}, allowing the apparent horizon to be geometrically identified. A \textit{future apparent horizon} is formally defined as the boundary of a closed spacelike hypersurface\footnote{In n--dimensional spacetime, spacelike hypersurfaces are submanifolds with $n-1$ dimensions whose boundary is $n-2$ dimensional surfaces. In 4--dimensional spacetime, a future apparent horizon is a 2-surface.} foliated by marginal surfaces. On these surfaces, null outgoing geodesic congruences are neither expanding nor contracting. This feature is best described by the scalar expansions $\theta_{l}, \theta_{n}$ corresponding to null future-directed outgoing and ingoing null geodesic congruences. Hence, on a future apparent horizon one gets $\theta_{l}=0, \quad \theta_{n}<0$. These horizons are defined in a \textit{quasi-local way}, meaning that an observer does not need to know the entire future of spacetime to localize them. Cosmological apparent horizons are usually \textit{past inner trapping horizons}, with conditions:
\begin{align}
    \theta_{n}=0, \quad \theta_{l}>0, \quad \mathscr{L}_{l}\theta_{n}>0.
    \label{conditions-past-trapping}
\end{align}
Here, $\mathscr{L}_{l}$ is a Lie derivative with respect to the outgoing future--directed null geodesics, with $l^{A}$. (See \cite{Faraoni:2015ula} references therein). In FLRW spacetimes, given by the spherically--symmetric metric \eqref{metric3}, the location of the apparent horizon can also be obtained by considering null hypersurfaces of constant areal radius, this results in setting $g^{RR}=0$. Therefore, the apparent horizon is located at:
\begin{align}
    R_{\rm{AH}}=\frac{1}{H}.
\end{align}
This expression gives the apparent-horizon areal radius. In terms of the comoving coordinate radius $r$, the apparent horizon is located at $r_{\rm AH}=1/\dot{a}$. In both cases, its location is model-dependent because it is determined by the Hubble parameter. In the de Sitter limit for $\Lambda$CDM, one obtains
\begin{align}
    R^{(dS)}_{\rm{AH}}=\sqrt{\frac{3}{\Lambda}}.
\end{align}
For the exponential $f(Q)$ model introduced in Eq.~\eqref{fexp}, it was shown
in Ref.~\cite{Vasquez:2025ywd} that the Hubble parameter satisfies the
transcendental equation
\begin{align}\label{HubExp}
3H^2+\Lambda e^{-(b\Lambda/6H^2)^{n}}\times\left[12H^2n(b\Lambda)^{n}(6H^2)^{-n-1}-1\right]=8\pi \rho,
\end{align}
where $\rho$ is the matter density. For $n=1$, the corresponding approximate
solution up to order $b^2$ is
\begin{gather}
H^2(z;b)= H^2_{0}\Biggl[\xi(z)-\frac{3\Omega^2_{\Lambda,0}}{2\xi(z)}b+\frac{\Omega^3_{\Lambda,0}(5\xi(z)-18\Omega_{\Lambda,0})}{8\xi^3(z)}b^2\Biggr],
\label{firstsolution}\\[0.5ex]
\begin{gathered}
H^2(z;b)= H^2_{0}\xi(z)\mathcal{F}(z;b;\Omega_{\Lambda,0}),\\
\mathcal{F}(z;b;\Omega_{\Lambda,0})=1-\frac{3\Omega^2_{\Lambda,0}}{2\xi^2(z)}b+\frac{\Omega^3_{\Lambda,0}(5\xi(z)-18\Omega_{\Lambda,0})}{8\xi^4(z)}b^2.
\end{gathered}
\end{gather}
This rewriting of $H(z)$, supposes an advantage, since $\mathcal{F}(z)=1$ for $\Lambda$CDM and also for the limit $z\rightarrow \infty$ (early universe), and at late times it is parametrized by $b$ in the form.
\begin{align}
\mathcal{F}(z;b;\Omega_{\Lambda,0})=1-\frac{3b}{2}-\frac{13b^2}{8}.
\end{align}
Here $\xi(z)=\Omega_{\Lambda,0}+\Omega_{m,0}(1+z)^3$ and $H_{0}$ is the Hubble parameter at the present epoch. Therefore, the apparent horizon radius for the exponential $f(Q)$ model is given by:
\begin{align}
    R_{\rm{AH}}=\frac{1}{H_{0}\sqrt{\xi(z)\mathcal{F}(z;b;\Omega_{\Lambda,0})}}.
\end{align}
Alternatively, the apparent-horizon radius may be evaluated using a numerical solution of the transcendental background equation. However, as shown in Ref.~\cite{Vasquez:2025ywd}, for the values of $b$ considered in the present analysis the numerical solution does not exhibit a significant discrepancy from the approximate solution in Eq.~\eqref{firstsolution}. We therefore, employ the approximate expression throughout the following analysis. Fig.~\ref{fig:apparent-horizon-radius} shows the redshift evolution of the dimensionless apparent-horizon radius for the observationally motivated best-fit values of $b$, together with the $\Lambda$CDM limit.

Figure~\ref{fig:apparent-horizon-radius} shows that, for comoving observers, the
apparent-horizon radius decreases toward higher redshifts and increases toward
lower redshifts. The geometric corrections that modify the gravitational
dynamics during cosmic expansion are most relevant at the present epoch and in
the late-time regime, and their magnitude is controlled by the parameter $b$. 

\subsection{Surface gravity and Kodama vector}
\begin{figure}[!ht]
  \centering
  \includegraphics[width=0.6\linewidth]{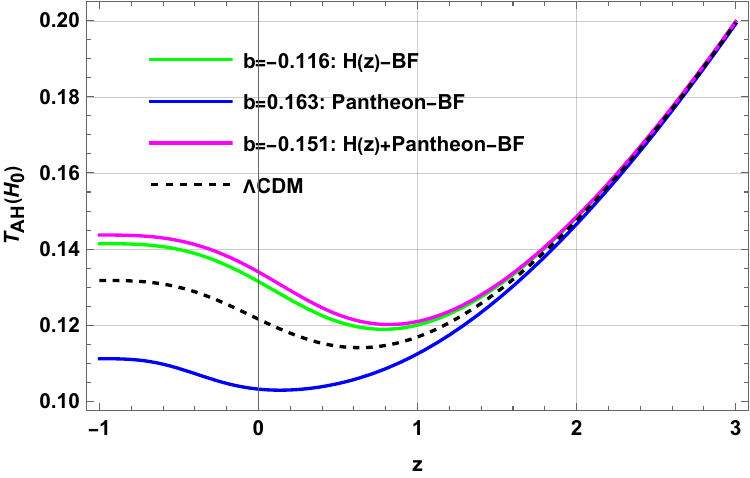}
  \caption{Redshift evolution of the Kodama--Hayward apparent-horizon temperature $T_{\rm AH}$ obtained from Eq.~\eqref{temperature--kodama} for the exponential $f(Q)$ model. The curves correspond to the best-fit values of the parameter $b$ reported in Ref.~\cite{Oliveros:2023bdv} and to the $\Lambda$CDM limit. The vertical axis is expressed in units of $H_0$.}
  \label{fig:apparent-horizon-temperature}
\end{figure}

Since the FLRW spacetime is dynamical, it does not admit a timelike Killing vector in general. Consequently, the usual definition of surface gravity based on the existence of a Killing horizon, as in stationary spacetimes such as Schwarzschild or Kerr, cannot be applied directly \cite{Faraoni:2015ula}. For spherically symmetric dynamical spacetimes, the Kodama vector provides a preferred time direction and plays a role analogous to that of a Killing vector \cite{Kodama:1979vn}. It is defined as:
\begin{equation}
    K^{A}=-\epsilon^{AB}\nabla_{B}R,
\end{equation}
where the indices $A,B$ run over the coordinates $(t,R)$ in the metric \eqref{metric2}. Here, $\epsilon^{AB}$ is the antisymmetric volume form for a 2-metric $h^{AB}$. Therefore, the Kodama vector is orthogonal to 2-spheres of symmetry in the surfaces $(t,R)$, that is $K^{A}\nabla_{A}R=0$ (by construction). The 2-induced metric casts the line element as
\begin{equation}
    ds^2=h^{AB}dx_{A}dx_{B}+R^2d\Omega^{2}_{(2)}.
\end{equation}
Here, $h^{AB}$ is given by the squared matrix, from metric \eqref{metric2}:
\begin{align}
h^{AB}= \frac{1}{\sqrt{|h|}}
\begin{pmatrix}
-1 & -HR \\
-HR & (1-H^2R^2)
\end{pmatrix},\qquad
h_{AB}= \frac{1}{\sqrt{|h|}}
\begin{pmatrix}
-(1-H^2R^2) & -HR \\
-HR & 1
\end{pmatrix}.
\end{align}
For which $\sqrt{|h|}=1$. The volume form $\epsilon^{AB}$ is thus given by the following expression:
\begin{equation}
    \epsilon^{AB}=\sqrt{|h|}h^{AC}h^{BD}(\nabla_{C}t\nabla_{D}R-\nabla_{C}R\nabla_{D}t).
\end{equation}
For the FLRW induced 2-metric \eqref{metric2} in $(t,R)$, the only nonzero components are given by $\epsilon^{tR}=-1, \epsilon^{Rt}=+1$ \cite{Faraoni:2015ula}. Therefore, the future-directed Kodama vector is:
\begin{align}
    K^{A}\partial_{A}=+\partial_{t}.
\end{align}
To confirm this, notice that the norm $K_{A}K^{A}=h_{tt}K^{t}K^{t}=-(1-H^2R^2)$ vanishes at the apparent horizon and $K^{A}$ becomes null on $\mathcal{AH}$\footnote{This notation will define the apparent horizon from now on.}. In dynamical spherically symmetric spacetimes, the surface gravity can be defined through the Kodama--Hayward expression \cite{Hayward:1997jp}:
\begin{equation}
\kappa_{\rm{Kodama}}=\frac{1}{2\sqrt{|h|}}\left.\partial_{\mu}\left(\sqrt{|h|}h^{\mu\nu}\partial_{\nu}R\right)\right.|_{\mathcal{AH}}.
\end{equation}
Therefore, the Kodama-Hayward surface gravity finally takes the usual form in flat FLRW geometry \cite{Rao:2024rhn, Nojiri:2024zdu}:
\begin{align}
\kappa_{\rm{AH}} = - \frac{1}{2H}(\dot{H}+2H^2)
\label{kodama-hayward-gravity}
\end{align}
Recalling the conditions for the apparent horizon to be a past-inner trapping horizon in Eq.~\eqref{conditions-past-trapping}, the Lie derivative along $l^{A}$ in FLRW spacetimes satisfies \cite{Faraoni:2015ula}
\begin{equation}
\mathcal{L}_{l}\theta_{n}=-2H\kappa_{\rm AH}>0.
\end{equation}
This condition is fulfilled since $\kappa_{\rm AH}$ in Eq.~\eqref{kodama-hayward-gravity} is negative. On the other hand, Hawking radiation in dynamical spacetimes has been extensively studied using tunneling methods, such as the Hamilton--Jacobi and Parikh--Wilczek approaches. The identification of a temperature for such horizons leads to the use of the Kodama--Hayward surface gravity \cite{Parikh:1999mf,Vanzo:2011wq,Faraoni:2015ula}. However, the sign of the temperature is determined by the sign of the Kodama--Hayward surface gravity at $\mathcal{AH}$. Since $\mathcal{AH}$ is a past-inner trapping horizon, positivity of the resulting Hawking temperature requires $T_{\rm AH}\propto-\kappa_{\rm AH}$ \cite{Rivadeneira-Caro:2025fcc}, the temperature of the FLRW apparent horizon is defined as:
\begin{align}
T_{\rm{AH}}=\frac{|\kappa_{\rm{AH}}|}{2\pi},
\end{align}
where $\hbar=1$. Therefore,
\begin{align}
T_{\rm{AH}}=\frac{1}{4\pi}\left(\frac{\dot{H}}{H}+2H\right), \quad   T_{\rm{AH}}=\frac{2H(z)-(1+z)H'}{4\pi}.
\end{align}
It is convenient to express the surface gravity in terms of $H(z)$ and rewrite $\dot{H}$ as a derivative w.r.t redshift.
In this way, the whole expression can be cast including contributions from the parameter $b$ of the exponential model. Here $H'(z)$ corresponds to the derivative  w.r.t $z$ of \eqref{firstsolution}, analytically given as:
\begin{align}
H'(z) &= \frac{1}{2}H_{0}\left[\frac{\xi'(z)}{\sqrt{\xi(z)}}\sqrt{\mathcal{F}(z;b)}+\frac{\mathcal{F}'(z;b)}{\sqrt{\mathcal{F}(z;b)}}\sqrt{\xi(z)}\right].
\end{align}
For completeness, it is also useful to recall the corresponding approximate
solution obtained in Ref.~\cite{Vasquez:2025ywd} for the quadratic exponential
case, $n=2$. This branch is not the one used in the numerical plots below, which
are based on the $n=1$ solution in Eq.~\eqref{firstsolution}, but it provides a
useful comparison because the leading correction depends on $b^2$ and is
therefore insensitive to the sign of the model parameter. For this solution, the
cosmological evolution does not differ significantly from the $\Lambda$CDM
behavior, as shown in Ref.~\cite{Vasquez:2025ywd}; hence, in what follows we
focus on the $n=1$ branch, where the dependence on $b$ produces a clearer
thermodynamic departure from the standard scenario. Then, the expression that
gives the Kodama-Hayward temperature at $\mathcal{AH}$ is:
\begin{align}
T_{\rm{AH}}(z;b)=\frac{H_{0}}{8\pi}\Biggl[4\sqrt{\xi(z)\mathcal{F}(z;b)}-3(1+z)^3\frac{\Omega_{m,0}}{\sqrt{\xi(z)}}\sqrt{\mathcal{F}(z;b)}-(1+z)\frac{\mathcal{F}'(z;b)}{\sqrt{\mathcal{F}(z;b)}}\sqrt{\xi(z)}\Biggr].
\label{temperature--kodama}
\end{align}
This expression allows us to evaluate how the exponential correction modifies the
thermal behavior of the apparent horizon along the cosmological evolution. In
Fig.~\ref{fig:apparent-horizon-temperature} we show the corresponding redshift
dependence of $T_{\rm AH}$ for the observationally motivated values of $b$ and
compare it with the $\Lambda$CDM limit, which is obtained by setting $b=0$ in
the expressions above.

Figure~\ref{fig:apparent-horizon-temperature} shows that the apparent-horizon
temperature remains positive over the redshift interval considered and follows a
smooth evolution for all the parameter choices. The curves tend to increase with
redshift, which is consistent with the fact that the apparent-horizon radius
$R_{\rm AH}=1/H$ decreases as the expansion rate grows toward the past. Since the Kodama--Hayward temperature is controlled by both $H(z)$ and its redshift derivative, the departure from the $\Lambda$CDM curve is mostly visible at low and intermediate redshifts, where the exponential correction in $F(z;b)$ contributes more appreciably to the background dynamics.

The effect of the parameter $b$ is therefore mainly to shift the normalization and slope of $T_{\rm AH}$ with respect to the $\Lambda$CDM behavior. The curves associated with negative values of $b$ remain close to the standard case, while the positive best-fit value produces a slightly higher temperature at late times.
At larger redshift the curves approach each other, indicating that the model recovers the expected early-time behavior when the exponential modification is suppressed. This reinforces the interpretation that the thermodynamic effects of the exponential $f(Q)$ sector are primarily late-time corrections to the apparent horizon temperature.

\section{Thermodynamics at the Apparent Horizon}
\label{sec:thermodynamics}

\subsection{Hayward's unified first law}
\label{subsec:first-law}

The first law for apparent horizons entails other difficulties, leading to a similar statement which has been dubbed \textit{unified first law} \cite{Hayward:1997jp,Hayward:1998ee}. In the first place, given a Schwarzschild-like metric as \eqref{metric3}, by direct comparison a quantity of \textit{mass} is given by:
\begin{align}
M_{\rm{FLRW}}=\frac{H^2\, R^3}{2}.
\label{FLRW-mass}
\end{align}
Evaluated at $R_{\rm{AH}}$, this corresponds to the mass-energy of the apparent horizon. Given by,
\begin{align}
M_{\rm{FLRW}}|_{\mathcal{AH}}\rightarrow M_{\rm{AH}}=\frac{R_{\rm{AH}}}{2}.
\label{FLRW-at-AH}
\end{align}
This same value of mass, identified by mere analogy with the Schwarzschild metric, can be obtained geometrically from the expression of the so-called \textbf{Misner-Sharp-Hernandez} mass \cite{Misner:1964je, Hernandez:1966zia}:
\begin{align}
M_{\rm{MSH}}=\frac{R}{2}(1-\nabla_{A}R\, \nabla^{A}R).
\end{align}
From which $\nabla_{A}R\nabla^{A}R=h^{AB}\nabla_{A}R\,\nabla_{B}R$. This expression is valid in GR for spherically symmetric spacetimes. Generalized Misner--Sharp energies have also been constructed in $f(R)$ and Gauss--Bonnet gravity \cite{Cai:2009qf,Maeda:2006pm}; however, to our knowledge, no corresponding generalization has yet been established for $f(Q)$ gravity in the coincident gauge. Since in GR, Eq.~\eqref{FLRW-mass} is equivalent to $\rho V$, with $V=4\pi R^3/3$, we may cast it effectively using Eq.~\eqref{eq:friedmann-1} as:
\begin{align}
M^{\rm{(eff)}}_{\rm{MSH}} &= \frac{R^3}{6}\left(Qf_{Q}-\frac{f}{2}\right)
\label{misner-sharp-hernandez}
\end{align}
When evaluated at $\mathcal{AH}$, this effective mass reduces to Eq.~\eqref{FLRW-at-AH} in the GR limit $f(Q)=Q$. A full covariant derivation of the generalized Misner--Sharp mass lies beyond the scope of the present work. The unified first law was given by Hayward in the following prescription: (i) define a work density function $W$ by the trace of the energy-momentum tensor:
\begin{align}
W &= -\frac{1}{2}T_{AB}h^{AB}, \\
& = -\frac{1}{2}[(\rho+p)u_{A}u_{B}+p\,h_{AB}]h^{AB}.
\end{align}
In the $(t,R)$ coordinates of the two-dimensional metric in Eq.~\eqref{metric2}, the comoving velocity has components $u^{A}=(1,HR)$ and $u_{A}=(-1,0)$, satisfying $u_{A}u^{A}=-1$. Therefore, the work density is
\begin{align}
W = \frac{1}{2}(\rho-p).
\end{align}
Using Eqs.~\eqref{eq:friedmann-1} and \eqref{eq:friedmann-2},  one obtains
\begin{align}
\label{density-fq}
\rho &= \frac{1}{8\pi}\left(Qf_Q-\frac{f}{2}\right), \\
p &= \frac{1}{8\pi}\left[\frac{f}{2}-Qf_Q-2\left(f_Q+2Qf_{QQ}\right)\dot{H}\right].
\end{align}
Therefore the work density function specified for a general $f(Q)$ model, results in:
\begin{equation}
    W=\frac{1}{16\pi}\left[2Qf_{Q}-f+2(f_{Q}+2Qf_{QQ})\dot{H} \right].
\end{equation}
Using the functional in Eq.~\eqref{fexp}, together with its first and second derivatives given in Eqs.~\eqref{eq:exponential-fq} and \eqref{eq:exponential-fqq}, and setting $n=1$. One then finds
\begin{equation}
    W=\frac{3H^2+\dot{H}}{8\pi}+\Lambda\,e^{-b\Lambda/6H^2}\Biggl[\frac{\Lambda b}{24\pi H^{2}}+\frac{b\,\dot{H}\Lambda(b\Lambda-9H^2)}{432\pi\, H^6}-\frac{1}{8\pi}\Biggr].
\end{equation}
Now, in the Hayward's prescription: (ii) a localized Bondi energy flux vector $\psi_{A}$ is defined across $\mathcal{AH}$ as:
\begin{align}
\nonumber
    \psi_{A} &=T^{B}_{A}\nabla_{B}R+W\nabla_{A}R,\\
        &= (\rho+p)\left[HR\,u_{A}+\frac{1}{2}\nabla_{A}R\right].
\end{align}
Using \eqref{eq:friedmann-2}, it results in:
\begin{align}
    \psi_{A}=-\frac{\dot{H}}{4\pi}\left( 2Qf_{QQ}+f_{Q}\right)\left[HR\,u_{A}+\frac{1}{2}\nabla_{A}R\right].
    \label{energy-flux}
\end{align}
Finally, for the particular model \eqref{fexp}, this can be casted as a combination of GR and modified $f(Q)$ gravity parts:
\begin{align}
    \psi_{A}=\psi^{(GR)}_{A}-\frac{b\,\dot{H}\, e^{-b\Lambda/6H^2}\Lambda^2(b\Lambda-9H^2)}{216\pi\, H^6}\times\left[HR\,u_{A}+\frac{1}{2}\nabla_{A}R\right].
\end{align}
Returning to Eq.~\eqref{energy-flux}, and since the volume enclosed by a sphere of areal radius \(R\) is \(V=4\pi R^3/3\), one has \(\nabla_A V=A\nabla_A R\), with \(A=4\pi R^2\). Using Eq.~\eqref{eq:friedmann-2}, the second term can also be written as $(\rho-W)\left.\nabla_A V\right|_{\mathcal{AH}}$, because \(\rho-W=(\rho+p)/2\). Hence, following Hayward: (iii) calculate the energy-supply vector at $\mathcal{AH}$, given as:
\begin{align}
    \left. A\psi_A\right|_{\mathcal{AH}}=-\frac{\dot H}{H^2}\left(2Qf_{QQ}+f_Q\right)u_{A}+(\rho-W)\left.\nabla_A V\right|_{\mathcal{AH}}.
    \label{eq:a-psi-ah-work}
\end{align}
Recognizing the form of the effective Misner-Sharp-Hernandez mass in Eq.~\eqref{misner-sharp-hernandez}, and using Eq.~\eqref{eq:friedmann-1}, one obtains
\begin{equation}
    \nabla_{A}M^{(\rm{eff})}_{\rm{MSH}}=\frac{V}{8\pi}\nabla_{A}\left(Qf_{Q}-\frac{f}{2}\right)+\frac{1}{8\pi}\left(Qf_{Q}-\frac{f}{2}\right)\nabla_{A}V,
\end{equation}
with,
\begin{align}
    \nabla_{A}Q=-\dot{Q}u_{A},\quad \nabla_{A}f_{Q}=-\dot{Q}f_{QQ}u_{A},\quad \nabla_{A}f=-\dot{Q}f_{Q}u_{A}.
\end{align}
Therefore,
\begin{equation}
    \nabla_{A}M^{\rm{(eff)}}_{\rm{MSH}}+\frac{\dot{H}}{H^2}(2Qf_{QQ}+f_{Q})u_{A}=\left.\rho\nabla_{A}V\right|_{\mathcal{AH}}.
\end{equation}
Hayward's unified first law is thus satisfied, independently of the $f(Q)$ functional at $\mathcal{AH}$:
\begin{equation}\label{haywardfirstlaw}
    \left(\nabla_{A}M^{\rm{(eff)}}_{\rm{MSH}}-A\psi_{A}-W\nabla_{A}V
    \right)\big{|_{\mathcal{AH}}}=0.
\end{equation}
This result shows that $f(Q)$ gravity in the coincident gauge yields equilibrium thermodynamics at the apparent horizon by employing an effective definition of the Misner-Sharp-Hernandez mass in Eq.~\eqref{misner-sharp-hernandez}. A similar result has been obtained in Ref.~\cite{Rao:2024rhn}, here it is demonstrated directly using Hayward's formulation of the unified first law. 

\subsection{Second law and entropy of cosmological apparent horizon}
The apparent horizon is a surface such that, $R=R_{\rm{AH}}$. A normal to this surface can be generally obtained as:
\begin{align}
n_{A} &= \nabla_{A}(R-R_{\rm{AH}}) \\
& = \nabla_{A}R+\frac{1}{H^2}\nabla_{A}H.
\end{align}
Naturally, the components are: $n_{R}=1, n_{t}=\frac{\dot{H}}{H^2}$. Hence, there must exist a vector $z^{A}$ such that $z^{A}n_{A}=0$, that is, a tangent to the surface \cite{Hayward:1997jp}. Given such condition, one gets:
\begin{align*}
\frac{\dot{H}}{H^2}\,z^{t}+z^{R}=0.
\end{align*}
Therefore, $z^{R}=-z^{t}\frac{\dot{H}}{H^2}$. Choosing the normalization $z^{t}=1$
\begin{align}
z^{A}\partial_{A} = -\frac{\dot{H}}{H^2}\partial_{R}+\partial_{t}.
\end{align}
As it is customary, the unified first law in the form \eqref{haywardfirstlaw}, can be projected along $z^{A}$ \cite{Hayward:1997jp}. Since the unified first law results as a rewriting of Einstein's field equations in spherically symmetric spacetimes, given the existence of a horizon, then a projection along $z^{A}$ renders the gravitational dynamics along apparent horizons. Generally, this results in:
\begin{align}
z^{A}\nabla_{A}M_{\rm{MSH}} = \dot{M}_{\rm{AH}}, \quad z^{A}\nabla_{A}V_{\rm{AH}}=\dot{V}_{\rm{AH}}.
\end{align}
Finally, the energy-supply vector is projected along the tangent $z^{A}$, thus giving the energy flow across $\mathcal{AH}$:
\begin{align*}
A\left.\psi_{A}z^{A}\right|_{\mathcal{AH}}=-R_{\rm{AH}}\kappa_{\rm{AH}}\frac{\dot{H}}{H^2}\,(2Qf_{QQ}+f_{Q}).
\end{align*}
Using $\dot{A}_{\rm{AH}}=-8\pi \frac{\dot{H}}{H^3}$, therefore one finds that this projection results in the time variation of the area as in the first law of black hole mechanics:
\begin{align}
A\left. \psi_{A}z^{A}\right|_{\mathcal{AH}}=\frac{\kappa_{\rm{AH}}}{8\pi}\dot{A}_{\rm{AH}}\,(2Qf_{QQ}+f_{Q}).
\end{align}
Thus, Hayward's unified first law \eqref{haywardfirstlaw} results equivalent to the \textbf{first law of thermodynamics} at an infinitesimal coordinate time interval $dt$:
\begin{align}
dM_{\rm{AH}}=A\left. \psi_{A}z^{A}\right|_{\mathcal{AH}}dt+WdV_{\rm{AH}}.
\end{align}
Following the work of Cai \textit{et al.}~\cite{Cai:2005ra}, the projected energy-supply term can be related to the entropy variation without introducing an entropy-production term:
\begin{align}
\label{entropy--clausius}
A\left.\psi_{A}z^{A}\right|_{\mathcal{AH}}dt=-T_{\rm{AH}}dS_{\rm{AH}}.
\end{align}
The minus sign reflects the orientation and heat flow conventions. With the future-directed horizon tangent adopted here, the FLRW apparent horizon is a past-inner trapping horizon and $\kappa_{\rm{AH}}<0$, where the physical temperature is defined as $T_{\rm{AH}}=|\kappa_{\rm{AH}}|/(2\pi)>0$. The quantity $A\psi_A z^A dt$ is therefore a signed energy-supply term rather than, by itself, the heat absorbed by the region enclosed by the horizon. Defining heat as positive when it enters this region gives
\begin{align}
  \delta Q_{\rm{AH}}
  \equiv -A\left.\psi_Az^A\right|_{\mathcal{AH}}dt
  =T_{\rm{AH}}dS_{\rm{AH}}.
\end{align}
Reversing the orientation of the horizon tangent reverses the flux sign, but leaves the physical content unchanged provided the convention is used consistently. The resulting apparent-horizon entropy is consistent with previous discussions of this subject \cite{Nojiri:2024zdu,Heydari:2025bbx,Nojiri:2025gkq}, and it shows an additional term in contrast with Noether--Wald entropy \cite{Momeni:2025rgk} which is usually proportional only to $f_{Q}$. It is worth analyzing the differences between Noether--Wald entropy, and entropy obtained from the UFL but it is out of the scope of this particular work. In particular, projecting Hayward's unified first law onto the apparent horizon yields an equilibrium thermodynamic description in which the horizon entropy is defined through the Clausius relation in Eq.~\eqref{entropy--clausius}. It is worth noting that Eq.~\eqref{density-fq} implies
\begin{align}
  \frac{\partial \rho}{\partial Q}
  =\frac{1}{16\pi}\left(2Qf_{QQ}+f_{Q}\right).
\end{align}
Therefore, the entropy differential at $\mathcal{AH}$ can be written as
\begin{align}
  dS_{\rm{AH}}
  =4\pi\frac{\partial \rho}{\partial Q}\,dA_{\rm{AH}}.
\end{align}
This relation expresses the infinitesimal entropy variation in terms of the response of the energy density to the non-metricity scalar.
The entropy of the apparent horizon is obtained by direct integration with respect to the trapping--horizon area:
\begin{align}
S(A)=\frac{1}{4}\int (2Qf_{QQ}+f_{Q})\, dA,
\end{align}
where the functional $f(Q)$ can be written in terms of the Bekenstein-Hawking area as:
\begin{align}
f(Q)\rightarrow f(A)=\frac{24\pi}{A}+2\Lambda\,e^{-b A\,\Lambda /24\pi}; \quad n=1.
\end{align}
Simplified, the integral is thus written as:
\begin{equation}
S(A)=\frac{1}{4}\int_{A_0}^A dA' \Biggl(1-e^{-bA'\,\Lambda/24\pi}\times\left[\frac{A'^2 b\,\Lambda^2}{96\pi^2}-\frac{A'^3b\,\Lambda^3}{3456\pi^3}\right]\Biggr).
\end{equation}
Integrating over the area results in corrections to the usual Bekenstein-Hawking entropy, \(S_{\rm{BH}}=A/4\). These corrections are of exponential nature, following the form of the functional as expected. The full integration leaves up to four extra terms:
\begin{equation}
S(A)=S_{\rm{BH}}-e^{-b A\,\Lambda/24\pi}\Biggl(\frac{72\pi}{b^2\Lambda}+\frac{3 A}{b}+\frac{\Lambda\,A^2}{16\pi}+\frac{\Lambda^2\,A^3 b}{576\pi^2}\Biggr)-S(A_0).
\label{eq:entropy-exponential-area}
\end{equation}
The structure of this result is reminiscent of the exponential corrections to black-hole entropy discussed in Ref.~\cite{Chatterjee:2020iuf}, where horizon microstate counting leads to corrections of the form $\exp[-A/(4\ell_p^2)]$ to the Bekenstein--Hawking area law. In the present case the exponential factor arises instead from the non-metricity function of the modified-gravity model and is controlled by the combination of cosmological parameters, $b\Lambda A/(24\pi)$. Thus, although the physical origin is different, Eq.~\eqref{eq:entropy-exponential-area} shares with Ref.~\cite{Chatterjee:2020iuf} the feature that the leading departure from the area law is exponentially suppressed in the horizon area rather than purely logarithmic or power-law. Implications of this exponential correction on Friedmann's equations and cosmological evolution has been explored, generally showing a higher incidence at low redshifts, and dynamics sensitive to the sign of introduced parameters (see Refs.~\cite{Pourhassan:2020yei, Rivadeneira-Caro:2025fcc}). These results are reminiscent of the cosmological dynamics generated by $f(Q)$ exponential model \cite{Vasquez:2025ywd}. 
Here, $S(A_0)$ denotes the integration constant obtained by evaluating the preceding expression at the lower limit $A_0$. Note that in the limit $b\rightarrow 0$, we obtain
\begin{equation}
  S(A)=\frac{A}{4}-\frac{A_0}{4}.
\end{equation}
This result becomes clearer by considering a series expansion around $b=0$:
\begin{equation}
S(A)=S_{\rm{BH}}-\frac{72\pi}{\Lambda b^2}-\frac{A^3\Lambda^2\,b}{1152\pi^2}+\frac{5A^4\Lambda^3\,b^2}{110592\pi^3}-S(A_{0})+\mathcal{O}(b^3).
\label{approximated-entropy}
\end{equation}
Since the second term is independent of \(A\), an identical contribution with the opposite sign appears in $S(A_0)$. These two terms therefore cancel each other, and the apparent divergence in the limit $b\rightarrow 0$ disappears. Finally, the integration constant may be fixed so that the standard Bekenstein--Hawking result, $S_{\rm{BH}}=A/4$, is recovered in the limit $b\rightarrow 0$. This avoids assigning direct physical significance to the formal choice $A_0=0$, which would correspond to a vanishing horizon area. In general, only the change in entropy between the initial and final states is physically relevant; thus, adding a constant value has no physical effect. As discussed in Ref.~\cite{Nojiri:2024zdu}, although different forms of horizon entropy may lead to different cosmological evolutions, they can share common properties. In particular, $S(A)$ is a monotonic function of the Bekenstein--Hawking entropy and satisfies $S(A)\rightarrow 0$ as $A_{\rm{BH}}\rightarrow 0$. For the present model, Eq.~\eqref{approximated-entropy} likewise gives $S(A)\rightarrow 0$ when $A=A_{\rm{BH}}\rightarrow 0$.

\begin{figure}[!ht]
  \centering
  \includegraphics[width=0.6\linewidth]{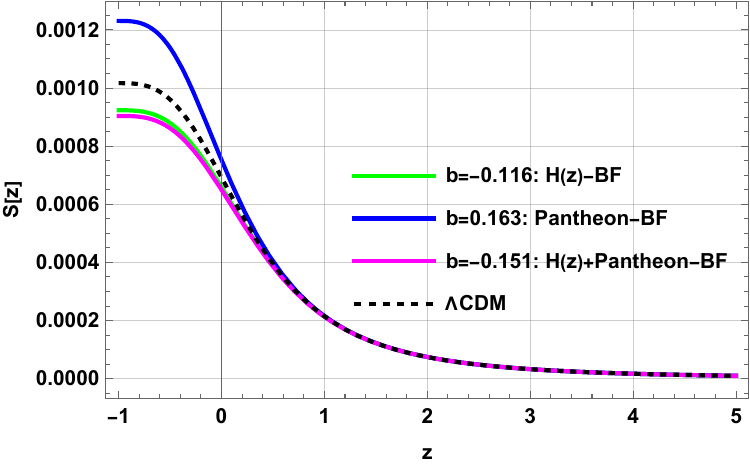}
  \caption{Redshift evolution of the normalized apparent-horizon entropy
  obtained from Eq.~\eqref{eq:entropy-exponential-area} for different values of
  $b$. The curves approach $S_{\rm AH}\to0$ at high redshift and converge toward
  the $\Lambda$CDM behavior in the early matter-dominated regime.}
  \label{fig:apparent-horizon-entropy}
\end{figure}

To display the entropy along the cosmological evolution, we evaluate
Eq.~\eqref{eq:entropy-exponential-area} using
$A(z)=4\pi/H^2(z;b)$, the approximate background solution in
Eq.~\eqref{firstsolution}, and the normalization $A_0=0$. The resulting
redshift dependence is shown in Fig.~\ref{fig:apparent-horizon-entropy} for the
observationally motivated values of $b$ and for the $\Lambda$CDM limit.

Figure~\ref{fig:apparent-horizon-entropy} shows that the apparent-horizon
entropy decreases as the redshift increases; equivalently, it grows as the
Universe evolves toward the present and future. The deviations from
$\Lambda$CDM are concentrated at low and negative redshift, where the
exponential correction is most relevant. The positive best-fit value of $b$
enhances the horizon entropy relative to $\Lambda$CDM, whereas the negative
values suppress it. At high redshift, all curves converge and approach zero as
the apparent-horizon area shrinks. Although this behavior is consistent with a
non-decreasing horizon entropy during the forward cosmic evolution, the entropy
of matter inside the horizon must also be included to establish the GSL.

\subsection{Generalized Second Law}
\label{subsec:gsl}
After obtaining the apparent-horizon entropy, the next step is to test whether
this entropy assignment is compatible with the GSL. In a
cosmological setting the relevant quantity is not only the horizon entropy, but
the total entropy obtained by adding the matter contribution inside the apparent
horizon. Therefore, the thermodynamic viability of the exponential model requires
that the total entropy does not decrease during the cosmic evolution.

In Ref.~\cite{Rao:2024rhn}, this condition was formulated for a multicomponent
$f(Q)$ universe by allowing each matter component to satisfy a continuity
equation with an interaction term $q_i$, while the total matter sector remains
conserved. Applying the first law to each component inside the apparent horizon,
the authors obtained the matter entropy production rate and combined it with the
horizon contribution,
\begin{equation}
  \dot S_A=-\frac{2\pi\dot H}{H^3}\left(f_Q+2Qf_{QQ}\right),
  \label{eq:rao-horizon-entropy-rate}
\end{equation}
to write the total entropy rate as
\begin{equation}
  \dot S_t=\dot S_m+\dot S_A.
  \label{eq:rao-total-entropy-rate}
\end{equation}
For a simplified thermal configuration in which all matter components share a
common temperature, the GSL reduces to a condition
proportional to $f_Q+2Qf_{QQ}$. If, in addition, the matter temperature is
identified with the horizon temperature, their result becomes
\begin{equation}
  \dot S_t=\frac{\dot H^2}{2H^4T}\left(f_Q+2Qf_{QQ}\right)\geq0,
  \label{eq:rao-gsl-equilibrium}
\end{equation}
For positive temperature this condition becomes
\begin{equation}
  f_Q+2Qf_{QQ}\geq 0.
  \label{eq:rao-gsl-viability} 
\end{equation}
Using the convention introduced in Eq.~\eqref{firstsolution}, we define
$\mathcal{E}(z;b)\equiv \xi(z)F(z;b;\Omega_{\Lambda,0})=H^2(z;b)/H_0^2$.
Then, the corresponding explicit expression used in the numerical analysis can
be written as
\begin{equation}
\dot S_t(z;b)=\frac{\pi(1+z)^2\left[\mathcal{E}'(z;b)\right]^2}{2H_0\,\mathcal{E}^4(z;b)\sqrt{\mathcal{E}(z;b)}}\times\frac{\Biggl[2\mathcal{E}^3(z;b)-b\Omega_{\Lambda,0}^2\exp\!\left(-\frac{b\Omega_{\Lambda,0}}{2\mathcal{E}(z;b)}\right)\times\left(3\mathcal{E}(z;b)-b\Omega_{\Lambda,0}\right)\Biggr]}{4\mathcal{E}(z;b)-(1+z)\mathcal{E}'(z;b)}.
\label{eq:explicit-total-entropy-rate}
\end{equation}
Here, the prime denotes differentiation with respect to the redshift $z$, and
$\Omega_{\Lambda,0}=1-\Omega_{m,0}$ for a spatially flat background.

\begin{figure*}[t]
  \centering
  \begin{minipage}{0.48\linewidth}
    \centering
    \includegraphics[width=\linewidth]{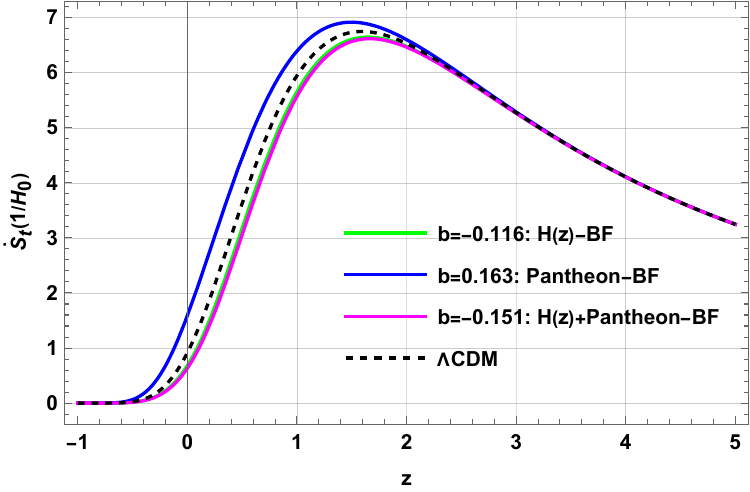}
    \vspace{0.5ex}
    \small (a) Total entropy production rate.
  \end{minipage}\hfill
  \begin{minipage}{0.48\linewidth}
    \centering
    \includegraphics[width=\linewidth]{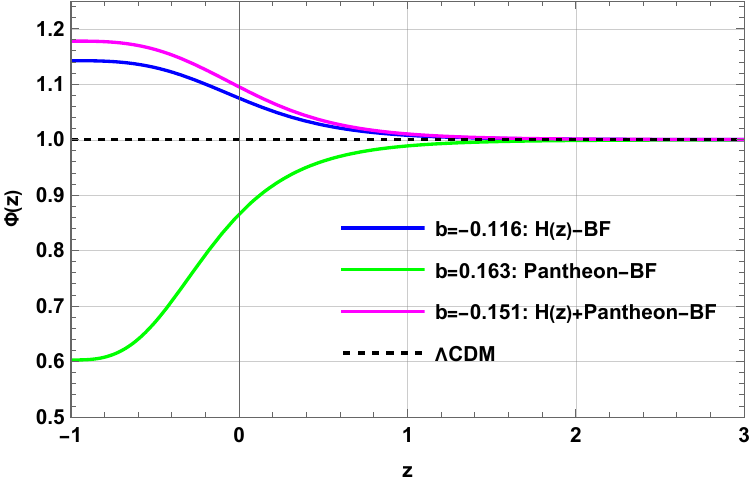}
    \vspace{0.5ex}
    \small (b) Viability function.
  \end{minipage}
  \caption{Thermodynamic viability of the present exponential $f(Q)$ model from the GSL. Left panel: evolution of the total entropy production rate $\dot S_t$ associated with Eq.~\eqref{eq:rao-gsl-equilibrium} for different values of the model parameter $b$, using the values reported in Ref.~\cite{Oliveros:2023bdv} and the approximate solution for $H(z;b)$ given by Eq.~\eqref{firstsolution} of the present work; the vertical axis is expressed in units of $1/H_0$. Right panel: redshift evolution of the viability function $\Phi(z)=f_Q+2Qf_{QQ}$ associated with Eq.~\eqref{eq:rao-gsl-viability}; the region $\Phi(z)\geq0$ corresponds to satisfaction of the GSL.}
  \label{fig:gsl-thermodynamic-viability}
\end{figure*}

Figure~\ref{fig:gsl-thermodynamic-viability} summarizes the thermodynamic
behavior associated with Eq.~\eqref{eq:rao-gsl-equilibrium} and the viability
condition that follows from it.

The behavior displayed in Fig.~\ref{fig:gsl-thermodynamic-viability} shows that
thermodynamic consistency is sensitive to the sign and magnitude of the
exponential correction. In the left panel, the total entropy production rate
remains non-negative for the parameter choices considered, indicating that the
the GSL is preserved along the corresponding cosmic evolution.
The curves also show a characteristic transient behavior: \(\dot S_t\) grows from
small values at low redshift, reaches a maximum at intermediate redshift, and
then decreases toward the high-redshift regime. This tendency reflects the fact
that the correction induced by the exponential sector is most relevant during the
transition between the late accelerated phase and the matter-dominated epoch,
whereas it becomes progressively suppressed when the model approaches the
standard cosmological behavior.

The right panel makes the same conclusion more transparent through the viability
function $\Phi(z)=f_Q+2Qf_{QQ}$. Since the prefactor multiplying $\Phi(z)$ in
Eq.~\eqref{eq:rao-gsl-equilibrium} is positive for positive temperature, the sign
of $\Phi(z)$ directly determines whether the GSL is
satisfied. The plotted trajectories approach the general-relativistic value
$\Phi(z)\simeq 1$ at sufficiently large redshift, as expected when the
exponential modification becomes negligible. At low redshift, however, the
curves separate according to the value of $b$, showing that the late-time
thermodynamic viability of the model imposes a non-trivial restriction on the
allowed parameter range. In particular, parameter choices for which
$\Phi(z)\geq0$ throughout the redshift interval considered correspond to a
non-decreasing total entropy, while any crossing into $\Phi(z)<0$ would signal
a violation of the GSL within the equilibrium-temperature
assumption used here. 

\begin{figure*}[t]
  \centering
  \begin{minipage}{0.48\linewidth}
    \centering
    \includegraphics[width=\linewidth]{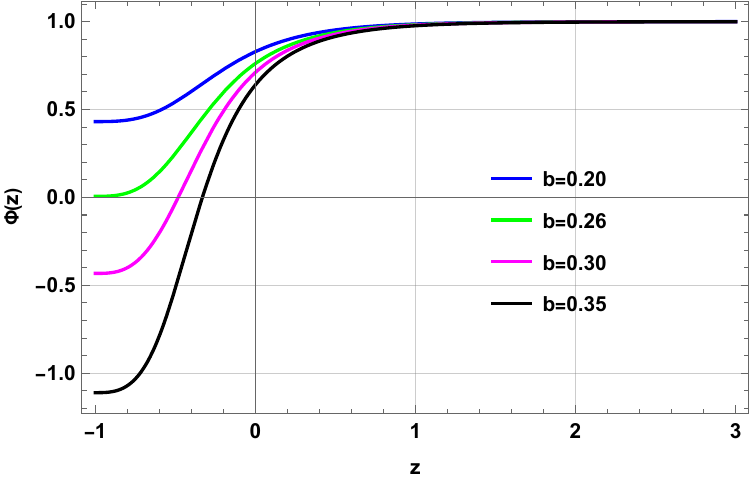}
    \vspace{0.5ex}
    \small (a) Representative one-dimensional slices.
  \end{minipage}\hfill
  \begin{minipage}{0.48\linewidth}
    \centering
    \includegraphics[width=\linewidth]{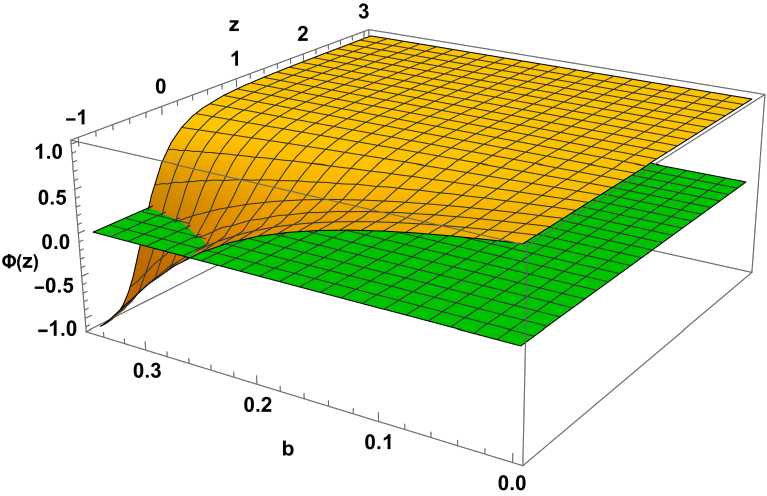}
    \vspace{0.5ex}
    \small (b) Two-parameter viability surface.
  \end{minipage}
  \caption{Thermodynamic viability of the exponential $f(Q)$ model for positive values of the parameter $b$. Left panel: redshift evolution of the viability function $\Phi(z)=f_Q+2Qf_{QQ}$ for selected values of $b$, with the horizontal line marking the threshold $\Phi(z)=0$. Right panel: three-dimensional representation of $\Phi(z,b)$ over the corresponding region of the $(z,b)$ plane, where the green plane denotes the boundary $\Phi=0$. Regions above this plane satisfy the GSL condition in Eq.~\eqref{eq:rao-gsl-viability}, whereas regions below it correspond to thermodynamically disfavored evolution.}
  \label{fig:gsl-viability-positive-b}
\end{figure*}

To make this constraint more explicit, it is useful to examine values of the
model parameter beyond the best-fit cases and to identify when the viability
function crosses the threshold $\Phi(z)=0$. This motivates the additional plot
shown in Fig.~\ref{fig:gsl-viability-positive-b}, where larger positive values
of $b$ are displayed separately.

Figure~\ref{fig:gsl-viability-positive-b} illustrates how the thermodynamic
viability condition becomes increasingly restrictive when the parameter $b$ is
moved toward larger positive values. Negative values of $b$ are not included in
this figure because, as shown in the right panel of
Fig.~\ref{fig:gsl-thermodynamic-viability}, the corresponding trajectories
remain in the region $\Phi(z)\geq0$ and therefore do not violate the GSL over
the redshift interval considered. In contrast, for positive
values of $b$, the function $\Phi(z)$ can become negative at low redshift. In
particular, the left panel shows that for values $b>0.26$ the GSL is violated in
the future-redshift region $z<0$, where $\Phi(z)<0$. This
behavior indicates that, within the equilibrium-temperature assumption used to
obtain Eq.~\eqref{eq:rao-gsl-equilibrium}, these parameter choices would violate
the GSL during the recent and near-future cosmological
evolution.

The right panel extends this conclusion by displaying the full surface
$\Phi(z,b)$ together with the plane $\Phi=0$. This representation makes clear
that the thermodynamically allowed domain corresponds to the portion of the
surface above the green plane, while the disfavored domain lies below it. The
surface bends downward as one moves toward larger positive $b$ and negative
redshift, showing that the violation is not an isolated feature of a single
curve but a continuous region in the $(z,b)$ parameter space. Hence, the boundary
where the surface intersects the plane $\Phi=0$ separates viable and non-viable
thermodynamic evolution.

The curves also show that the violation is mainly a late-time effect. As the
redshift increases, all trajectories rise and approach positive values close to
the general-relativistic regime, reflecting the suppression of the exponential
correction at earlier epochs. The depth of the negative region depends strongly
on $b$: larger positive values drive $\Phi(z)$ farther below zero and postpone
the recovery of the viable regime to higher redshift. Therefore, the combined
one-dimensional and three-dimensional plots provide a direct thermodynamic upper
constraint on positive values of $b$, since the requirement $\Phi(z)\geq0$ over
the relevant redshift interval excludes those parameter choices for which the
surface crosses below the zero threshold.

The restriction to the coincident-gauge branch deserves a final clarification
from the thermodynamic perspective. The analysis in Ref.~\cite{Rao:2024rhn} shows that, for genuinely non-trivial cosmological
connection branches, the entropy variation cannot in general be integrated as a
state function determined only by the apparent-horizon radius and temperature.
Instead, it depends on the detailed cosmic evolution and contains an additional
entropy-production contribution, thereby characterizing a non-equilibrium
thermodynamic description. Since the present work is specifically formulated
within equilibrium horizon thermodynamics and applies the corresponding GSL
criterion, including such branches would require a separate non-equilibrium
treatment rather than a direct extension of the current analysis. Our use of
the coincident gauge is therefore a deliberate methodological restriction and
does not imply that non-trivial connections are physically inadmissible. Their
cosmological stability and phenomenology remain important separate questions;
related results can be found in Ref.~\cite{Guzman:2024cwa}.


\section{Discussion and Conclusions}
\label{sec:conclusion}

In this work we investigated the thermodynamics of the apparent horizon in the
exponential $f(Q)$ gravity model defined in Eq.~\eqref{fexp}. The analysis was
performed for a spatially flat FLRW background in the coincident gauge, with
particular emphasis on the $n=1$ branch and on the approximate cosmological
solution for $H(z;b)$ up to second order in the parameter $b$. This framework
allowed us to connect the background evolution of the model with the geometric
properties of the apparent horizon and with the corresponding thermodynamic
relations.

The apparent-horizon radius remains determined by $R_{\rm AH}=1/H$, but its
redshift evolution acquires a characteristic dependence on $b$ through the
modified Hubble function. For the observationally motivated values considered
here, the deviations from $\Lambda$CDM are concentrated mainly at low and
intermediate redshifts, whereas the curves approach the standard behavior toward
the high-redshift regime. The Kodama--Hayward temperature exhibits the same
general tendency: it remains positive and evolves smoothly over the interval
studied, with the exponential correction primarily modifying its late-time
normalization and slope. These results support the interpretation that the
thermodynamic effects of the exponential non-metricity sector are predominantly
late-time phenomena.

Using an effective form for Misner--Sharp--Hernandez mass in $f(Q)$, the work density, and the energy-supply
vector, we showed directly that Hayward's unified first law is satisfied at the
apparent horizon. In the coincident-gauge FLRW branch considered in this work,
the horizon dynamics can therefore be described within an equilibrium
thermodynamic framework. Projecting the unified first law along the tangent to
the horizon leads to an entropy differential proportional to
$f_Q+2Qf_{QQ}$. For the exponential model, the resulting entropy contains
exponentially suppressed corrections to the Bekenstein--Hawking area law. The
functional structure is therefore reminiscent of the exponential entropy
correction reported in Ref.~\cite{Chatterjee:2020iuf}. The resemblance is
formal rather than physical: in that work the correction originates from the
counting of quantum states localized at a black-hole horizon, whereas here it is
generated by the exponential non-metricity sector and its dependence on the
cosmological parameters. The integration constant can be chosen consistently so
that the standard result $S_{\rm BH}=A/4$ is recovered in the limit $b\to0$,
while the apparent divergent constant terms cancel between the upper and lower
limits of the entropy integral.

The GSL provides an additional and independent viability
criterion. Under the assumption that the matter components share a common
temperature equal to the apparent-horizon temperature, its sign is controlled by
the function $\Phi=f_Q+2Qf_{QQ}$. The best-fit parameter values examined in
Fig.~\ref{fig:gsl-thermodynamic-viability} yield a non-negative total entropy
production rate over the redshift range considered. However, the extended
positive-$b$ analysis in Fig.~\ref{fig:gsl-viability-positive-b} shows that the
condition becomes restrictive in the future domain: values $b>0.26$ can drive
$\Phi$ below zero for $z<0$. Within the equilibrium assumptions adopted here,
this identifies a thermodynamically disfavored region and provides an upper
constraint on sufficiently large positive values of the exponential parameter.

Several extensions deserve further study. The present analysis could be
generalized to arbitrary values of the exponent $n$. It would also be useful to
relax the common-temperature assumption and formulate the entropy balance in a
genuinely non-equilibrium setting, allowing for energy exchange among the matter
sectors and the horizon. Further directions include spatially curved FLRW
geometries, non-trivial flat and torsionless connection branches, and the
inclusion of cosmological perturbations. Finally, thermodynamic viability could
be combined with observational constraints from cosmic-chronometer $H(z)$
measurements, Pantheon+ Type Ia supernovae, DESI DR2 baryon-acoustic-oscillation
data, and CMB observations from Planck
\cite{Brout:2022vxf,DESI:2025zgx,Planck:2018vyg}. Growth-sector tests based on
redshift-space-distortion measurements of $f\sigma_8(z)$ and weak-lensing and
galaxy-clustering data from DES Year 3 would provide complementary constraints
on the perturbative behavior of the model
\cite{Kazantzidis:2018rnb,DES:2021wwk}.

\bibliographystyle{unsrt}
{\raggedright
\bibliography{references}

@article{Jimenez:2019qfb,
  author = {Jimenez, Jose Beltran and Heisenberg, Lavinia and Koivisto, Tomi S. and Pekar, Simon},
  title = {Cosmology in {$f(Q)$} geometry},
  journal = {Phys. Rev. D},
  volume = {101},
  pages = {103507},
  year = {2020},
  doi = {10.1103/PhysRevD.101.103507},
  eprint = {1906.10027},
  archivePrefix = {arXiv},
  primaryClass = {gr-qc}
}

@article{DAmbrosio:2023cwr,
  author = {D'Ambrosio, Fabio and Heisenberg, Lavinia and Zentarra, Stefan},
  title = {Hamiltonian Analysis of {$f(Q)$} Gravity and the Failure of the Dirac--Bergmann Algorithm for Teleparallel Theories of Gravity},
  journal = {Fortsch. Phys.},
  volume = {71},
  pages = {2300185},
  year = {2023},
  doi = {10.1002/prop.202300185},
  eprint = {2308.02250},
  archivePrefix = {arXiv},
  primaryClass = {gr-qc}
}

@article{Tomonari:2023wcs,
  author = {Tomonari, Kyosuke and Bahamonde, Sebastian},
  title = {Dirac--Bergmann analysis and degrees of freedom of coincident {$f(Q)$} gravity},
  journal = {Eur. Phys. J. C},
  volume = {84},
  pages = {349},
  year = {2024},
  doi = {10.1140/epjc/s10052-024-12677-x},
  eprint = {2308.06469},
  archivePrefix = {arXiv},
  primaryClass = {gr-qc}
}

@article{Gomes:2023tur,
  author = {Gomes, Debora Aguiar and Jimenez, Jose Beltran and Cano, Alejandro Jimenez and Koivisto, Tomi S.},
  title = {Pathological Character of Modifications to Coincident General Relativity: Cosmological Strong Coupling and Ghosts in {$f(Q)$} Theories},
  journal = {Phys. Rev. Lett.},
  volume = {132},
  pages = {141401},
  year = {2024},
  doi = {10.1103/PhysRevLett.132.141401},
  eprint = {2311.04201},
  archivePrefix = {arXiv},
  primaryClass = {gr-qc}
}

@misc{BeltranJimenez:2026wja,
    author = "Beltr{\'a}n Jim{\'e}nez, Jose and Jim{\'e}nez Cano, Alejandro and Koivisto, Tomi S.",
    title = "{Physical nonviability of $f(\mathbb{Q})$ in the scalar-tensor representation}",
    eprint = "2606.30036",
    note = {arXiv:2606.30036 [gr-qc]},
    archivePrefix = "arXiv",
    primaryClass = "gr-qc",
    month = "6",
    year = "2026"
}

@article{Heisenberg:2023lgi,
  author = {Heisenberg, Lavinia and Hohmann, Manuel and Kuhn, Simon},
  title = {Cosmological teleparallel perturbations},
  journal = {JCAP},
  volume = {03},
  pages = {063},
  year = {2024},
  doi = {10.1088/1475-7516/2024/03/063},
  eprint = {2311.05495},
  archivePrefix = {arXiv},
  primaryClass = {gr-qc}
}

@article{Linder:2009jz,
  author = {Linder, Eric V.},
  title = {Exponential Gravity},
  journal = {Phys. Rev. D},
  volume = {80},
  pages = {123528},
  year = {2009},
  doi = {10.1103/PhysRevD.80.123528},
  eprint = {0905.2962},
  archivePrefix = {arXiv},
  primaryClass = {astro-ph.CO}
}

@article{Elizalde:2010ts,
  author = {Elizalde, E. and Nojiri, S. and Odintsov, S. D. and Sebastiani, L. and Zerbini, S.},
  title = {Non-singular exponential gravity: A simple theory for early- and late-time accelerated expansion},
  journal = {Phys. Rev. D},
  volume = {83},
  pages = {086006},
  year = {2011},
  doi = {10.1103/PhysRevD.83.086006},
  eprint = {1012.2280},
  archivePrefix = {arXiv},
  primaryClass = {hep-th}
}

@article{Li:2013xea,
  author = {Li, Jung-Tsung and Lee, Chung-Chi and Geng, Chao-Qiang},
  title = {Einstein Static Universe in Exponential {$f(T)$} Gravity},
  journal = {Eur. Phys. J. C},
  volume = {73},
  pages = {2315},
  year = {2013},
  doi = {10.1140/epjc/s10052-013-2315-z},
  eprint = {1302.2688},
  archivePrefix = {arXiv},
  primaryClass = {gr-qc}
}

@article{Pradhan:2025gsl,
  author = {Pradhan, Anirudh and Husain, A. and Zeyauddin, M. and Shekh, S. H.},
  title = {Generalized Second Law and Thermodynamical Aspects of {$f(Q,\mathcal{T})$} Gravity},
  journal = {Annals Phys.},
  volume = {490},
  pages = {170498},
  year = {2026},
  doi = {10.1016/j.aop.2026.170498},
  eprint = {2510.12863},
  archivePrefix = {arXiv},
  primaryClass = {gr-qc}
}

@article{Pervaiz:2026npb,
  author = {Pervaiz, Nageen and Azhar, Nadeem and Li, Jin-Zeng and Jawad, Abdul and Myrzakulov, Nurgissa and others},
  title = {Late-time cosmic acceleration in torsion-free {$f(Q)$} gravity: Dynamical and thermodynamical analysis},
  journal = {Nucl. Phys. B},
  volume = {1029},
  pages = {117517},
  year = {2026},
  doi = {10.1016/j.nuclphysb.2026.117517}
}

@article{Weinberg:1988cp,
  author = {Weinberg, Steven},
  title = {The cosmological constant problem},
  journal = {Rev. Mod. Phys.},
  volume = {61},
  pages = {1--23},
  year = {1989},
  doi = {10.1103/RevModPhys.61.1}
}

@article{Sahni:1999gb,
  author = {Sahni, Varun and Starobinsky, Alexei A.},
  title = {The case for a positive cosmological {$\Lambda$}-term},
  journal = {Int. J. Mod. Phys. D},
  volume = {9},
  pages = {373--444},
  year = {2000},
  doi = {10.1142/S0218271800000542},
  eprint = {astro-ph/9904398},
  archivePrefix = {arXiv}
}

@article{Peebles:2002gy,
  author = {Peebles, P. J. E. and Ratra, Bharat},
  title = {The cosmological constant and dark energy},
  journal = {Rev. Mod. Phys.},
  volume = {75},
  pages = {559--606},
  year = {2003},
  doi = {10.1103/RevModPhys.75.559},
  eprint = {astro-ph/0207347},
  archivePrefix = {arXiv}
}

@article{Copeland:2006wr,
  author = {Copeland, Edmund J. and Sami, M. and Tsujikawa, Shinji},
  title = {Dynamics of dark energy},
  journal = {Int. J. Mod. Phys. D},
  volume = {15},
  pages = {1753--1936},
  year = {2006},
  doi = {10.1142/S021827180600942X},
  eprint = {hep-th/0603057},
  archivePrefix = {arXiv}
}

@article{Riess:1998cb,
  author = {Riess, Adam G. and others},
  collaboration = {Supernova Search Team},
  title = {Observational evidence from supernovae for an accelerating universe and a cosmological constant},
  journal = {Astron. J.},
  volume = {116},
  pages = {1009--1038},
  year = {1998},
  doi = {10.1086/300499},
  eprint = {astro-ph/9805201},
  archivePrefix = {arXiv}
}

@article{Perlmutter:1998np,
  author = {Perlmutter, S. and others},
  collaboration = {Supernova Cosmology Project},
  title = {Measurements of {$\Omega$} and {$\Lambda$} from 42 high-redshift supernovae},
  journal = {Astrophys. J.},
  volume = {517},
  pages = {565--586},
  year = {1999},
  doi = {10.1086/307221},
  eprint = {astro-ph/9812133},
  archivePrefix = {arXiv}
}

@article{Planck:2018vyg,
  author = {Aghanim, N. and others},
  collaboration = {Planck},
  title = {Planck 2018 results. {VI}. Cosmological parameters},
  journal = {Astron. Astrophys.},
  volume = {641},
  pages = {A6},
  year = {2020},
  doi = {10.1051/0004-6361/201833910},
  eprint = {1807.06209},
  archivePrefix = {arXiv},
  primaryClass = {astro-ph.CO}
}

@article{Riess:2021jrx,
  author = {Riess, Adam G. and others},
  title = {A comprehensive measurement of the local value of the {Hubble} constant with 1 km s$^{-1}$ Mpc$^{-1}$ uncertainty from the {Hubble Space Telescope} and the {SH0ES} team},
  journal = {Astrophys. J. Lett.},
  volume = {934},
  number = {1},
  pages = {L7},
  year = {2022},
  doi = {10.3847/2041-8213/ac5c5b},
  eprint = {2112.04510},
  archivePrefix = {arXiv},
  primaryClass = {astro-ph.CO}
}

@article{Jimenez:2018bfs,
  author = {Jim\'enez, Jose Beltr\'an and Heisenberg, Lavinia and Koivisto, Tomi},
  title = {Coincident General Relativity},
  journal = {Phys. Rev. D},
  volume = {98},
  number = {4},
  pages = {044048},
  year = {2018},
  doi = {10.1103/PhysRevD.98.044048},
  eprint = {1710.03116},
  archivePrefix = {arXiv},
  primaryClass = {gr-qc}
}

@article{Jimenez:2019ovq,
  author = {Jim\'enez, Jose Beltr\'an and Heisenberg, Lavinia and Koivisto, Tomi S.},
  title = {The Geometrical Trinity of Gravity},
  journal = {Universe},
  volume = {5},
  number = {7},
  pages = {173},
  year = {2019},
  doi = {10.3390/universe5070173},
  eprint = {1903.06830},
  archivePrefix = {arXiv},
  primaryClass = {hep-th}
}

@article{Lazkoz:2019sjl,
  author = {Lazkoz, Ruth and Lobo, Francisco S. N. and Ortiz-Ba\~nos, Mar\'ia and Salzano, Vincenzo},
  title = {Observational constraints of {$f(Q)$} gravity},
  journal = {Phys. Rev. D},
  volume = {100},
  number = {10},
  pages = {104027},
  year = {2019},
  doi = {10.1103/PhysRevD.100.104027},
  eprint = {1907.13219},
  archivePrefix = {arXiv},
  primaryClass = {gr-qc}
}

@article{Anagnostopoulos:2021ydo,
  author = {Anagnostopoulos, Fotios K. and Basilakos, Spyros and Saridakis, Emmanuel N.},
  title = {First evidence that non-metricity {$f(Q)$} gravity could challenge {$\Lambda$CDM}},
  journal = {Phys. Lett. B},
  volume = {822},
  pages = {136634},
  year = {2021},
  doi = {10.1016/j.physletb.2021.136634},
  eprint = {2104.15123},
  archivePrefix = {arXiv},
  primaryClass = {gr-qc}
}

@article{Heisenberg:2023lru,
  author = {Heisenberg, Lavinia},
  title = {Review on {$f(Q)$} gravity},
  journal = {Phys. Rept.},
  volume = {1066},
  pages = {1--78},
  year = {2024},
  doi = {10.1016/j.physrep.2024.02.001},
  eprint = {2309.15958},
  archivePrefix = {arXiv},
  primaryClass = {gr-qc}
}

@article{Bekenstein:1973ur,
  author = {Bekenstein, Jacob D.},
  title = {Black holes and entropy},
  journal = {Phys. Rev. D},
  volume = {7},
  pages = {2333--2346},
  year = {1973},
  doi = {10.1103/PhysRevD.7.2333}
}

@article{Hawking:1975vcx,
  author = {Hawking, S. W.},
  title = {Particle creation by black holes},
  journal = {Commun. Math. Phys.},
  volume = {43},
  pages = {199--220},
  year = {1975},
  doi = {10.1007/BF02345020}
}

@article{Bardeen:1973gs,
    author = "Bardeen, James M. and Carter, B. and Hawking, S. W.",
    title = "{The Four laws of black hole mechanics}",
    doi = "10.1007/BF01645742",
    journal = "Commun. Math. Phys.",
    volume = "31",
    pages = "161--170",
    year = "1973"
}

@article{Jacobson:1995ab,
  author = {Jacobson, Ted},
  title = {Thermodynamics of spacetime: The Einstein equation of state},
  journal = {Phys. Rev. Lett.},
  volume = {75},
  pages = {1260--1263},
  year = {1995},
  doi = {10.1103/PhysRevLett.75.1260},
  eprint = {gr-qc/9504004},
  archivePrefix = {arXiv}
}

@article{Cai:2005ra,
  author = {Cai, Rong-Gen and Kim, Sang Pyo},
  title = {First law of thermodynamics and Friedmann equations of Friedmann-Robertson-Walker universe},
  journal = {JHEP},
  volume = {02},
  pages = {050},
  year = {2005},
  doi = {10.1088/1126-6708/2005/02/050},
  eprint = {hep-th/0501055},
  archivePrefix = {arXiv}
}

@article{Akbar:2006kj,
  author = {Akbar, M. and Cai, Rong-Gen},
  title = {Thermodynamic behavior of field equations for {$f(R)$} gravity},
  journal = {Phys. Lett. B},
  volume = {648},
  pages = {243--248},
  year = {2007},
  doi = {10.1016/j.physletb.2007.03.005},
  eprint = {gr-qc/0612089},
  archivePrefix = {arXiv}
}

@article{Bamba:2009gq,
  author = {Bamba, Kazuharu and Geng, Chao-Qiang},
  title = {Thermodynamics in {$f(R)$} gravity in the Palatini formalism},
  journal = {JCAP},
  volume = {06},
  pages = {014},
  year = {2010},
  doi = {10.1088/1475-7516/2010/06/014},
  eprint = {1005.5234},
  archivePrefix = {arXiv},
  primaryClass = {gr-qc}
}

@article{Bamba:2012rv,
  author = {Bamba, Kazuharu and Geng, Chao-Qiang and Lee, Chung-Chi and Luo, Ling-Wei},
  title = {Equation of state for dark energy in {$f(T)$} gravity},
  journal = {JCAP},
  volume = {01},
  pages = {021},
  year = {2011},
  doi = {10.1088/1475-7516/2011/01/021},
  eprint = {1011.0508},
  archivePrefix = {arXiv},
  primaryClass = {astro-ph.CO}
}

@article{Rao:2024rhn,
  author = {Rao, Haomin and Liu, Chunhui and Geng, Chao-Qiang},
  title = {Thermodynamic of the {$f(Q)$} universe},
  journal = {Eur. Phys. J. C},
  volume = {84},
  pages = {1317},
  year = {2024},
  doi = {10.1140/epjc/s10052-024-13711-8},
  eprint = {2406.09036},
  archivePrefix = {arXiv},
  primaryClass = {gr-qc}
}

@article{Nojiri:2025fiu,
    author = "Nojiri, Shin'ichi and Odintsov, Sergei D.",
    title = "{The correspondence of generalised entropic cosmology theory with F(T) and F(Q) modified gravity and gravitational waves}",
    eprint = "2502.15272",
    archivePrefix = "arXiv",
    primaryClass = "gr-qc",
    reportNumber = "KEK-TH-2690, KEK-Cosmo-0373",
    doi = "10.1016/j.dark.2025.101899",
    journal = "Phys. Dark Univ.",
    volume = "48",
    pages = "101899",
    year = "2025"
}

@article{Khyllep:2022spx,
  author = {Khyllep, Wompherdeiki and Dutta, Jibitesh and Saridakis, Emmanuel N. and Yesmakhanova, Kuralay},
  title = {Cosmology in {$f(Q)$} gravity: A unified dynamical system analysis at background and perturbation levels},
  journal = {Phys. Rev. D},
  volume = {107},
  number = {4},
  pages = {044022},
  year = {2023},
  doi = {10.1103/PhysRevD.107.044022},
  eprint = {2207.02610},
  archivePrefix = {arXiv},
  primaryClass = {gr-qc}
}

@article{Narawade:2023jfl,
  author = {Narawade, S. A. and Singh, Shashank P. and Mishra, B.},
  title = {Accelerating cosmological models in {$f(Q)$} gravity and the phase space analysis},
  journal = {Phys. Dark Univ.},
  volume = {42},
  pages = {101282},
  year = {2023},
  doi = {10.1016/j.dark.2023.101282},
  eprint = {2303.06427},
  archivePrefix = {arXiv},
  primaryClass = {gr-qc}
}

@article{Sokoliuk:2023eha,
  author = {Sokoliuk, Oleksii and Arora, Simran and Praharaj, Subhrat and Baransky, Alexander and Sahoo, P. K.},
  title = {On the impact of {$f(Q)$} gravity on the large scale structure},
  journal = {Mon. Not. Roy. Astron. Soc.},
  volume = {522},
  number = {1},
  pages = {252--267},
  year = {2023},
  doi = {10.1093/mnras/stad968},
  eprint = {2303.17341},
  archivePrefix = {arXiv},
  primaryClass = {astro-ph.CO}
}

@article{Mhamdi:2024glb,
  author = {Mhamdi, Dalale and Bargach, Farida and Dahmani, Safae and Bouali, Amine and Ouali, Taoufik},
  title = {Constraints on power law and exponential models in {$f(Q)$} gravity},
  journal = {Phys. Lett. B},
  volume = {859},
  pages = {139113},
  year = {2024},
  doi = {10.1016/j.physletb.2024.139113},
  eprint = {2410.10480},
  archivePrefix = {arXiv},
  primaryClass = {gr-qc}
}

@article{Odintsov:2017qif,
    author = "Odintsov, Sergei D. and S{\'a}ez-Chill{\'o}n G{\'o}mez, Diego and Sharov, German S.",
    title = "{Is exponential gravity a viable description for the whole cosmological history?}",
    eprint = "1709.06800",
    archivePrefix = "arXiv",
    primaryClass = "gr-qc",
    doi = "10.1140/epjc/s10052-017-5419-z",
    journal = "Eur. Phys. J. C",
    volume = "77",
    number = "12",
    pages = "862",
    year = "2017"
}

@article{Oliveros:2023bdv,
  author = {Oliveros, A. and Acero, Mario A.},
  title = {Cosmological dynamics and observational constraints on a viable {$f(Q)$} non-metric gravity model},
  journal = {Int. J. Mod. Phys. D},
  volume = {33},
  number = {01},
  pages = {2450004},
  year = {2024},
  doi = {10.1142/S0218271824500044},
  eprint = {2311.01857},
  archivePrefix = {arXiv},
  primaryClass = {astro-ph.CO}
}

@article{Vasquez:2025ywd,
  author = {Vasquez, Ivan R. and Oliveros, A.},
  title = {Analysis of the cosmological evolution parameters, energy conditions, and linear matter perturbations of an exponential-type model in {$f(Q)$} gravity},
  journal = {Gen. Rel. Grav.},
  volume = {57},
  number = {4},
  pages = {67},
  year = {2025},
  doi = {10.1007/s10714-025-03403-3},
  eprint = {2501.12585},
  archivePrefix = {arXiv},
  primaryClass = {gr-qc}
}

@article{Vasquez:2025tbg,
  author = {Vasquez, Ivan R. and Oliveros, A.},
  title = {Phase space analysis of an exponential model in {$f(Q)$} gravity including linear dark-sector interactions},
  journal = {Eur. Phys. J. C},
  volume = {86},
  pages = {104},
  year = {2026},
  doi = {10.1140/epjc/s10052-026-15313-y},
  eprint = {2510.12020},
  archivePrefix = {arXiv},
  primaryClass = {gr-qc}
}

@article{Ganz:2025ydt,
    author = "Ganz, Alexander and Spinelli, Marco",
    title = "{Ghost instabilities and strong coupling in quadratic non-metricity theories}",
    journal = "JCAP",
    volume = "05",
    pages = "104",
    year = "2026",
    doi = "10.1088/1475-7516/2026/05/104",
    eprint = "2511.19101",
    archivePrefix = "arXiv",
    primaryClass = "gr-qc",
}

@article{DiCriscienzo:2007pcr,
    author = "Di Criscienzo, R. and Nadalini, M. and Vanzo, L. and Zerbini, S. and Zoccatelli, G.",
    title = "{On the Hawking radiation as tunneling for a class of dynamical black holes}",
    eprint = "0707.4425",
    archivePrefix = "arXiv",
    primaryClass = "hep-th",
    doi = "10.1016/j.physletb.2007.10.005",
    journal = "Phys. Lett. B",
    volume = "657",
    pages = "107--111",
    year = "2007"
}

@article{Nojiri:2010wj,
  author = {Nojiri, Shin'ichi and Odintsov, Sergei D.},
  title = {Unified cosmic history in modified gravity: From {$F(R)$} theory to Lorentz non-invariant models},
  journal = {Phys. Rept.},
  volume = {505},
  pages = {59--144},
  year = {2011},
  doi = {10.1016/j.physrep.2011.04.001},
  eprint = {1011.0544},
  archivePrefix = {arXiv},
  primaryClass = {gr-qc}
}

@article{Clifton:2011jh,
  author = {Clifton, Timothy and Ferreira, Pedro G. and Padilla, Antonio and Skordis, Constantinos},
  title = {Modified gravity and cosmology},
  journal = {Phys. Rept.},
  volume = {513},
  pages = {1--189},
  year = {2012},
  doi = {10.1016/j.physrep.2012.01.001},
  eprint = {1106.2476},
  archivePrefix = {arXiv},
  primaryClass = {astro-ph.CO}
}

@article{Joyce:2016vqv,
  author = {Joyce, Austin and Lombriser, Lucas and Schmidt, Fabian},
  title = {Dark energy versus modified gravity},
  journal = {Ann. Rev. Nucl. Part. Sci.},
  volume = {66},
  pages = {95--122},
  year = {2016},
  doi = {10.1146/annurev-nucl-102115-044553},
  eprint = {1601.06133},
  archivePrefix = {arXiv},
  primaryClass = {astro-ph.CO}
}

@article{Guzman:2024cwa,
    author = {Guzm{\'a}n, Mar{\'\i}a-Jos{\'e} and J{\"a}rv, Laur and Pati, Laxmipriya},
    title = "{Exploring the stability of f(Q) cosmology near general relativity limit with different connections}",
    eprint = "2406.11621",
    archivePrefix = "arXiv",
    primaryClass = "gr-qc",
    doi = "10.1103/PhysRevD.110.124013",
    journal = "Phys. Rev. D",
    volume = "110",
    number = "12",
    pages = "124013",
    year = "2024"
}

@article{Nolan:1998xs,
    author = "Nolan, Brien C.",
    title = "{A Point mass in an isotropic universe: Existence, uniqueness and basic properties}",
    eprint = "gr-qc/9805041",
    archivePrefix = "arXiv",
    doi = "10.1103/PhysRevD.58.064006",
    journal = "Phys. Rev. D",
    volume = "58",
    pages = "064006",
    year = "1998"
}

@book{Faraoni:2015ula,
    author = "Faraoni, Valerio",
    title = "{Cosmological and Black Hole Apparent Horizons}",
    doi = "10.1007/978-3-319-19240-6",
    isbn = "978-3-319-19239-0, 978-3-319-19240-6",
    publisher = "Springer",
    volume = "907",
    year = "2015"
}

@article{Hayward:1997jp,
    author = "Hayward, Sean A.",
    title = "{Unified first law of black hole dynamics and relativistic thermodynamics}",
    eprint = "gr-qc/9710089",
    archivePrefix = "arXiv",
    doi = "10.1088/0264-9381/15/10/017",
    journal = "Class. Quant. Grav.",
    volume = "15",
    pages = "3147--3162",
    year = "1998"
}

@article{Kodama:1979vn,
    author = "Kodama, Hideo",
    title = "{Conserved Energy Flux for the Spherically Symmetric System and the Back Reaction Problem in the Black Hole Evaporation}",
    reportNumber = "KUNS-506",
    doi = "10.1143/PTP.63.1217",
    journal = "Prog. Theor. Phys.",
    volume = "63",
    pages = "1217",
    year = "1980"
}

@article{Parikh:1999mf,
    author = "Parikh, Maulik K. and Wilczek, Frank",
    title = "{Hawking radiation as tunneling}",
    eprint = "hep-th/9907001",
    archivePrefix = "arXiv",
    reportNumber = "PUPT-1775, SPIN-1998-12, IASSNS-HEP-98-22",
    doi = "10.1103/PhysRevLett.85.5042",
    journal = "Phys. Rev. Lett.",
    volume = "85",
    pages = "5042--5045",
    year = "2000"
}

@article{Vanzo:2011wq,
    author = "Vanzo, L. and Acquaviva, G. and Di Criscienzo, R.",
    title = "{Tunnelling Methods and Hawking's radiation: achievements and prospects}",
    eprint = "1106.4153",
    archivePrefix = "arXiv",
    primaryClass = "gr-qc",
    doi = "10.1088/0264-9381/28/18/183001",
    journal = "Class. Quant. Grav.",
    volume = "28",
    pages = "183001",
    year = "2011"
}

@article{Hayward:1998ee,
    author = "Hayward, Sean A. and Mukohyama, Shinji and Ashworth, M. C.",
    title = "{Dynamic black hole entropy}",
    eprint = "gr-qc/9810006",
    archivePrefix = "arXiv",
    doi = "10.1016/S0375-9601(99)00225-X",
    journal = "Phys. Lett. A",
    volume = "256",
    pages = "347--350",
    year = "1999"
}

@article{Rivadeneira-Caro:2025fcc,
    author = "Rivadeneira-Caro, Rodrigo and Saavedra, Joel F. and Tello-Ortiz, Francisco",
    title = "{Cosmological FLRW phase transitions under exponential corrected entropy}",
    journal = "Fortschr. Phys.",
    volume = "74",
    number = "3",
    pages = "e70063",
    doi = "10.1002/prop.70063",
    eprint = "2509.11919",
    archivePrefix = "arXiv",
    primaryClass = "hep-th",
    year = "2026"
}

@article{Nojiri:2024zdu,
    author = "Nojiri, Shin'ichi and Odintsov, Sergei D. and Paul, Tanmoy",
    title = "{Different Aspects of Entropic Cosmology}",
    eprint = "2409.01090",
    archivePrefix = "arXiv",
    primaryClass = "gr-qc",
    doi = "10.3390/universe10090352",
    journal = "Universe",
    volume = "10",
    number = "9",
    pages = "352",
    year = "2024"
}

@article{Cai:2009qf,
    author = "Cai, Rong-Gen and Cao, Li-Ming and Hu, Ya-Peng and Ohta, Nobuyoshi",
    title = "{Generalized Misner-Sharp Energy in f(R) Gravity}",
    eprint = "0910.2387",
    archivePrefix = "arXiv",
    primaryClass = "hep-th",
    reportNumber = "KU-TP-036",
    doi = "10.1103/PhysRevD.80.104016",
    journal = "Phys. Rev. D",
    volume = "80",
    pages = "104016",
    year = "2009"
}

@article{Maeda:2006pm,
    author = "Maeda, Hideki",
    title = "{Final fate of spherically symmetric gravitational collapse of a dust cloud in Einstein-Gauss-Bonnet gravity}",
    eprint = "gr-qc/0602109",
    archivePrefix = "arXiv",
    doi = "10.1103/PhysRevD.73.104004",
    journal = "Phys. Rev. D",
    volume = "73",
    pages = "104004",
    year = "2006"
}

@article{Misner:1964je,
    author = "Misner, Charles W. and Sharp, David H.",
    title = "{Relativistic equations for adiabatic, spherically symmetric gravitational collapse}",
    doi = "10.1103/PhysRev.136.B571",
    journal = "Phys. Rev.",
    volume = "136",
    pages = "B571--B576",
    year = "1964"
}

@article{Hernandez:1966zia,
    author = "Hernandez, Walter C. and Misner, Charles W.",
    title = "{Observer Time as a Coordinate in Relativistic Spherical Hydrodynamics}",
    doi = "10.1086/148525",
    journal = "Astrophys. J.",
    volume = "143",
    pages = "452",
    year = "1966"
}

@article{DESI:2024mwx,
  author = {Adame, A. G. and others},
  collaboration = {DESI},
  title = {{DESI} 2024 {VI}: Cosmological constraints from the measurements of baryon acoustic oscillations},
  journal = {JCAP},
  volume = {02},
  pages = {021},
  year = {2025},
  doi = {10.1088/1475-7516/2025/02/021},
  eprint = {2404.03002},
  archivePrefix = {arXiv},
  primaryClass = {astro-ph.CO}
}

@article{DESI:2025zgx,
  author = {Adame, A. G. and others},
  collaboration = {DESI},
  title = {{DESI DR2} results. {II}. Measurements of baryon acoustic oscillations and cosmological constraints},
  journal = {Phys. Rev. D},
  volume = {112},
  number={8},
  pages = {083515},
  year = {2025},
  doi = {10.1103/PhysRevD.112.083515},
  eprint = {2503.14738},
  archivePrefix = {arXiv},
  primaryClass = {astro-ph.CO}
}

@article{Momeni:2025rgk,
    author = "Momeni, Davood and Myrzakulov, Ratbay",
    title = "{Wald Entropy in Extended Modified Myrzakulov Gravity Theories: $f(R, T, Q, R_{\mu \nu }T^{\mu \nu }, R_{\mu \nu }Q^{\mu \nu }, \dots )$}",
    eprint = "2511.03509",
    archivePrefix = "arXiv",
    primaryClass = "gr-qc",
    doi = "10.1007/s10773-025-06143-x",
    journal = "Int. J. Theor. Phys.",
    volume = "64",
    number = "10",
    pages = "268",
    year = "2025"
}

@article{Nojiri:2025gkq,
    author = "Nojiri, Shin'ichi and Odintsov, Sergei D. and Paul, Tanmoy and SenGupta, Soumitra",
    title = "{Modified gravity as entropic cosmology}",
    journal = "Universe",
    volume = "12",
    number = "5",
    pages = "126",
    doi = "10.3390/universe12050126",
    eprint = "2503.19056",
    archivePrefix = "arXiv",
    primaryClass = "gr-qc",
    year = "2026"
}

@article{Heydari:2025bbx,
    author = "Heydari, Soma and Askari, Parastoo and Karami, Kayoomars",
    title = "{Revisited apparent horizon entropy and GSL in modified gravity}",
    eprint = "2512.17861",
    archivePrefix = "arXiv",
    primaryClass = "gr-qc",
    doi = "10.1140/epjc/s10052-026-15832-8",
    journal = "Eur. Phys. J. C",
    volume = "86",
    number = "6",
    pages = "603",
    year = "2026"
}

@article{Chatterjee:2020iuf,
  author = {Chatterjee, Ayan and Ghosh, Amit},
  title = {Exponential Corrections to Black Hole Entropy},
  journal = {Phys. Rev. Lett.},
  volume = {125},
  number = {4},
  pages = {041302},
  year = {2020},
  doi = {10.1103/PhysRevLett.125.041302},
  eprint = {2007.15401},
  archivePrefix = {arXiv},
  primaryClass = {gr-qc}
}

@article{Pourhassan:2020yei,
    author = "Pourhassan, Behnam",
    title = "{Exponential corrected thermodynamics of black holes}",
    eprint = "2010.03946",
    archivePrefix = "arXiv",
    primaryClass = "gr-qc",
    doi = "10.1088/1742-5468/ac0f6a",
    journal = "J. Stat. Mech.",
    volume = "2107",
    pages = "073102",
    year = "2021"
}

@article{Brout:2022vxf,
  author = {Brout, Dillon and others},
  title = {The {Pantheon+} Analysis: Cosmological Constraints},
  journal = {Astrophys. J.},
  volume = {938},
  number = {2},
  pages = {110},
  year = {2022},
  doi = {10.3847/1538-4357/ac8e04},
  eprint = {2202.04077},
  archivePrefix = {arXiv},
  primaryClass = {astro-ph.CO}
}

@article{Kazantzidis:2018rnb,
  author = {Kazantzidis, Lavrentios and Perivolaropoulos, Leandros},
  title = {Evolution of the $f\sigma_8$ tension with the {Planck15}/$\Lambda${CDM} determination and implications for modified gravity theories},
  journal = {Phys. Rev. D},
  volume = {97},
  number = {10},
  pages = {103503},
  year = {2018},
  doi = {10.1103/PhysRevD.97.103503},
  eprint = {1803.01337},
  archivePrefix = {arXiv},
  primaryClass = {astro-ph.CO}
}

@article{DES:2021wwk,
  author = {Abbott, T. M. C. and others},
  collaboration = {DES},
  title = {Dark Energy Survey Year 3 Results: Cosmological Constraints from Galaxy Clustering and Weak Lensing},
  journal = {Phys. Rev. D},
  volume = {105},
  number = {2},
  pages = {023520},
  year = {2022},
  doi = {10.1103/PhysRevD.105.023520},
  eprint = {2105.13549},
  archivePrefix = {arXiv},
  primaryClass = {astro-ph.CO}
}
\par}

\end{document}